\documentclass[12pt]{article}

\AtBeginDocument{
	\addtocontents{toc}{\normalsize}%\footnotesize
}
\pdfoutput=1

\usepackage{amsmath,amssymb,amsfonts,amsthm,amscd,mathrsfs}
\usepackage{esvect}
\usepackage{xcolor}
\definecolor{darkblue}{rgb}{0.1,0.1,.7}
\usepackage[]{version}
\usepackage[]{graphicx}
\usepackage[]{latexsym}
\usepackage{geometry}
\usepackage[all,cmtip]{xy}

\usepackage[margin=10pt,font=small,labelfont=bf]{caption}
\usepackage{ifthen}
\usepackage{soul}
\usepackage{tikz}
\usepackage{array,mathrsfs,amsfonts,yfonts,dsfont,bbm,colonequals}
\usepackage[nodisplayskipstretch]{setspace}
\usepackage{dsfont}
\usepackage{cite}
\usepackage{xspace}
\usepackage{empheq}
\usepackage{extarrows}

\usepackage{xcolor}
\usepackage[colorlinks, linkcolor=darkblue, citecolor=darkblue, urlcolor=darkblue, linktocpage]{hyperref} 

\usepackage{subcaption}
\usepackage[utf8]{inputenc}
\usepackage{setspace}
\usepackage{float, graphicx}

\numberwithin{equation}{section}

\newcommand{\aF}{\alpha_F}
\newcommand{\af}{\tilde \alpha}
\newcommand{\Fi}{\Psi}

\newcommand{\cO}{\mathcal O}

\newcommand{\reef}[1]{(\ref{#1})}
\newcommand{\be}{\begin{equation}}
\newcommand{\ee}{\end{equation}}
\newcommand{\bea}{\begin{eqnarray}}
\newcommand{\eea}{\end{eqnarray}}
\newcommand{\ba}{\begin{equation}\begin{aligned}}
\newcommand{\ea}{\end{aligned}\end{equation}}
\newcommand{\ttext}[1]{\mbox{\tiny #1}}

\newcommand{\ud}{\mathrm d}

\newcommand{\Df}{{\Delta_\phi}}

\def\g{\gamma}

\def\m{\mu}

\def\a{\alpha}
\def\e{\epsilon}

\def\b{\beta}

\def\D{\Delta}
\def\G{\Gamma}
\def\l{\lambda}

\def\la{\langle}

\def\ml{{\mathcal{L}}}
\def\G{\Gamma}

\def\la{\label}
\def\rf{\eqref}
\usepackage{cancel}
\def\p{\phi}

\newcommand{\ha}{\tfrac{1}{2}}

\newcommand{\no}{\nonumber}
\newcommand{\h}{h}

\begin{document}
	
	\vspace*{-.6in} \thispagestyle{empty}
	\begin{flushright}
		%CERN PH-TH/2015-200\\
		%LPTENS/18/18
	\end{flushright}
	%%
	%% Title
	%%
	\vspace{1cm} {\Large
		\begin{center}
			{\bf Bootstrapping bulk locality\\ \vspace{0.2cm}
  Part I:  Sum rules for AdS form factors}\\
	\end{center}}
	\vspace{1cm}
	%%%
	%% Authors
	%%%
	\begin{center}
		{\bf Nat Levine$^{a,b}$   and    Miguel F.~Paulos$^a$}\\[1cm] 
		{
			\small
			{\em ${}^a$Laboratoire de Physique, \qquad ${}^b$Institut Philippe Meyer, \\ \'Ecole Normale Sup\'erieure, \\
   Universit{\'e} PSL, CNRS, Sorbonne Universit{\'e}, Universit{\'e} Paris Cit{\'e}, \\
24 rue Lhomond, F-75005 Paris, France}

			\normalsize
		}
		
		%\vspace{1cm}\today
	\end{center}
	
	\begin{center}
		{  \texttt{nat.levine@phys.ens.fr  , \  miguel.paulos@ens.fr} 
		}
		\\
		%\vspace{1cm}\today
	\end{center}
	
	\vspace{8mm}
	
	\begin{abstract}
 \vspace{2mm}
    The problem of constructing local bulk observables from boundary CFT data is of paramount importance in holography. In this work, we begin addressing this question from a modern bootstrap perspective. Our main tool is the boundary operator expansion (BOE), which holds for any QFT in AdS. Following Kabat and Lifschytz, we argue that the BOE is strongly constrained by demanding locality of correlators involving bulk fields. Focusing on `AdS form factors' of one bulk and two boundary insertions, 
    we reformulate these locality constraints as a complete set of sum rules on the BOE data. We show that these sum rules lead to 
    a manifestly local representation 
    of form factors
    in terms of `local blocks'. The sum rules are valid non-perturbatively, but are especially well-adapted for perturbative computations in AdS where they allow us to bootstrap the BOE data in a systematic fashion. Finally, in the flat space limit, we show that the AdS form factor reduces to an ordinary QFT form factor. We provide a phase shift formula for it in terms of the BOE and CFT data. In two dimensions, this formula  makes manifest Watson's equations for integrable form factors under certain extremality assumptions on the CFT. We discuss the eventual modifications of our formalism to account for dressed operators in AdS.
    
	\end{abstract}
	\vspace{2in}

	\newpage
	
	{
		\setlength{\parskip}{0.05in}
		\tableofcontents
		\renewcommand{\baselinestretch}{1.0}\normalsize
	}
	
	%\newpage
	
	\setlength{\parskip}{0.1in}
 \setlength{\abovedisplayskip}{15pt}
 \setlength{\belowdisplayskip}{15pt}
 \setlength{\belowdisplayshortskip}{15pt}
 \setlength{\abovedisplayshortskip}{15pt}
 
	% \newpage
	
	%\input{Sections/Introduction}
	\bigskip \bigskip
	\section{Introduction}
	Suppose you are given a CFT: how would you know whether it supports an (approximately) local description in AdS? This is usually taken to mean that the theory has an equivalent formulation in terms of a reasonable, usually weakly coupled, effective field theory in AdS. Key requirements on the CFT are then some kind of large $N$ expansion (responsible for weak coupling) and a large gap in the spectrum of `single-trace' operators (so that there is a finite number of fields in the description) \cite{Heemskerk:2009pn}. Over the years, a number of works have shown that these assumptions do indeed lead to effective theories in AdS in line with expectations (see e.g. \cite{Fitzpatrick:2012cg,Camanho:2014apa,Afkhami-Jeddi:2016ntf,Caron-Huot:2021enk}).
 
    There is, however, a different but related way of thinking about this problem. If such an AdS description exists, then there should be a canonical and systematic way to construct operators in the CFT that behave as weakly interacting local operators propagating inside an asymptotically AdS space. 
    This way of thinking about the problem is an old one, going back almost to the beginning of AdS/CFT, with work by Bena \cite{Bena:1999jv} and then major contributions from Hamilton, Kabat, Lifschytz and Lowe \cite{Hamilton:2005ju,
Hamilton:2006az,
Hamilton:2006fh,
Hamilton:2007wj,
Kabat:2011rz,
Kabat:2012hp,
Kabat:2012av,
Kabat:2013wga,
Kabat:2014kfa,
Kabat:2015swa,
Kabat:2016zzr,
Foit:2019nsr,
Kabat:2020nvj}. These works essentially reverse-engineer AdS perturbative computations to understand the structure of couplings of bulk fields to CFT operators. One of the key results \cite{Hamilton:2006az,Almheiri:2014lwa} is the construction of free AdS fields in terms of smeared CFT operators in a finite causal domain whose size shrinks as fields approach the boundary. Another is that, as interactions are switched on, bulk fields generically couple to infinitely many CFT operators \cite{Kabat:2011rz}. This may lead one to worry whether, at finite coupling, this expansion is still well-defined and, in particular, whether it commutes with the smearing.
    
    In this work, we will take a bootstrap approach to the problem of constructing local AdS operators from the CFT. The basic idea was first articulated by Kabat and Lifschytz \cite{Kabat:2016zzr}: impose locality of correlators which involve bulk fields in order to constrain their couplings to boundary operators. Our work brings two new, key ideas to 
    this problem. Firstly, instead of using smearing kernels, we will rely on the basic fact that any QFT in AdS has  a version of the state-operator correspondence, called the \textit{boundary operator expansion} (BOE). The BOE allows any state --- in particular, ones created by local bulk insertions $\Psi$ --- to be expanded as a sum of boundary operators:
    % a bulk operator $\Psi$ can be expressed as
    % it takes the form
    \unskip\footnote{This is a generalization of a similar statement for boundary CFT \cite{Diehl:1981jg,Cardy:1991tv}.}
     schematically,
    \ba
    \Psi(x,u)\sim \sum_{\cO} \mu^\Psi_{\cO}\, u^{\Delta_\cO}\, \cO(x) 
    \ea   
    in AdS coordinates $u,x^\mu$. 
    The BOE allows correlators involving bulk fields to be expressed as convergent sums of boundary quantities, even non-perturbatively. Our approach will thus be similar to, but more general than, the approaches for bulk reconstruction set out in \cite{Verlinde:2015qfa,Anand:2017dav}.    
    Our second new element 
    is the development of machinery  
    to translate 
    locality of bulk correlators 
    into
    rigorous constraints, taking the form of 
    convergent sum rules on the BOE data. These are generally valid for any QFT in AdS, and they turn out to be especially well adapted for perturbative computations, as we show in a number of examples. This is closely related to similar technology developed in recent years for extracting constraints from various kinds of bootstrap equations \cite{
Mazac:2016qev,
Mazac:2019shk,
Mazac:2018mdx,
Mazac:2018ycv,
Mazac:2018biw,
Kaviraj:2018tfd,
Paulos:2019fkw,
Paulos:2019gtx,
Paulos:2020zxx,
Ghosh:2021ruh,
Caron-Huot:2020adz,
Bissi:2019kkx,
Giombi:2020xah,
Penedones:2019tng,
Carmi:2020ekr,
Gopakumar:2021dvg,
Ferrero:2019luz,
Gopakumar:2018xqi,
Dey:2016mcs,
Gopakumar:2016cpb,
Gopakumar:2016wkt
}.
Concretely, by considering correlators of $\Psi$ with two boundary insertions we find
    \ba
    \langle \Psi \, \cO_1 \,  \cO_2\rangle \  \mbox{is local} \qquad \Leftrightarrow \qquad \sum_{\cO\in \cO_1\times \cO_2} \mu^{\Psi}_{\cO} \, \lambda^{\cO_1 \cO_2}_{\cO} \, \theta^{12}_n(\Delta_{\cO})=0\, \quad n=1,2\ldots \ .
    \ea
    These are not formal expressions, but well-defined, absolutely convergent sum rules. The functions $\theta_n^{12}(\Delta)$ are theory-independent and can be determined in terms of simple ratios of gamma functions. By considering different pairs of operators $\cO_1$ and $\cO_2$, these equations not only constrain the BOE coefficients $\mu^\Psi_\cO$ but also the CFT OPE data $\lambda^{\cO_1\cO_2}_{\cO}$.

    In this article, we will assume  exact locality of bulk fields and address the question of how to express them in terms of boundary data. 
    We focus on UV-complete QFTs placed in AdS with arbitrary choices of curvature couplings and boundary conditions. Such theories naturally lead to families of $d$ dimensional CFTs  living on the boundary of spacetime, labeled by dimensionless quantities such as the mass gap in 
    AdS units.  Our setup applies, in particular, to boundary conformal field theory (BCFT): it is nothing but the special case where the QFT in AdS is also a CFT (and so the setup is Weyl-equivalent to a CFT in a flat half-space). 
    
    We will be ignoring things such as gravitational or gauge symmetries in the bulk (i.e.\ considering only gauge-invariant local operators if a gauge symmetry is present). 
    One may wonder about the fate of our approach for genuinely holographic (i.e.\ gravitational) AdS theories, where it should not be possible to define exactly local observables. As we will argue in Section \ref{Disc}, the technology  
    developed in this work should still be relevant for operators charged under gauge symmetries, or in gravitational theories. We simply expect 
    to have to add a certain source term to the sum rules developed here. Thus, even though locality ultimately fails, we expect it to do so in a sufficiently controlled way that, at least under certain assumptions, our formalism is still useful.

    We will also explore the flat space limit of observables involving boundary and bulk insertions, where the AdS radius $R$ is sent to infinity. We will argue that, in gapped QFTs, the flat space limit of these quantities directly reproduces flat space form factors upon a suitable analytic continuation:
    % . Specifically, we will show:
    %
    \ba
    \langle \Psi \, \cO_1 \, \cO_2\rangle \underset{\substack{R\to \infty\\\mbox {\tiny continuation}}}{\longrightarrow} \langle 0| \Psi |k_1,k_2\rangle \ .
    \ea
    In particular, we will derive a formula for the latter in terms of the BOE data. Writing
    \ba
    \langle 0|\Psi(x)|k_1,k_2\rangle=e^{i k x} \mathcal F^{\Psi}_{12}(s),\qquad s=-(k_1+k_2)^2=E^2 \ ,
    \ea
    we find
    \ba
    \mathcal F^{\Psi}_{12}(s)=\lim_{R\to \infty}\sum_{\cO\in \cO_1\times \cO_2}e^{-i\pi \frac{\Delta-\D_1-\D_2}{2}}\, \left(\frac{\mu^\Psi_{\cO}\lambda^{\cO_1\cO_2}_{\cO}}{\hat c^{12}_{\Delta_{\cO}}}\right)\, \mathcal N_R(\Delta_\cO,E R) \ ,
    \ea
    where $\mathcal N_R$ is a gaussian of variance $\sim 1/R$ centered at $\Delta_{\cO}=E R$. 
    This formula is very similar to one that was derived for the flat space S-matrix in terms of the boundary CFT data \cite{QFTinAdS,Komatsu:2020sag,Cordova:2022pbl,vanRees:2022itk}. We will show that it leads to correct results in a number of examples.

\bigskip

    The structure of this work is as follows. In Section \ref{SAdS}, we introduce the locality problem, after explaining the relevant kinematics of AdS form factors and the BOE. In Section \ref{Sloc}, we recast the locality condition as a dispersion relation, and equivalently as a manifestly local decomposition of the form factor in terms of `local blocks' (analogous to Polyakov blocks in the Polyakov bootstrap). In Section \ref{SR}, locality is formulated as a complete list of functional sum rules.

    In Section \ref{WA}, we apply our sum rules to bootstrap form factors for free scalar field theories in AdS, and their perturbation by $\Phi^4$ type interactions. We identify the local blocks as `exchange' Witten diagrams and thus obtain explicit expressions for the functionals defined above. In Section \ref{loccft}, we demonstrate in the context of free theories that the BOE coefficients may be `eliminated' to give an infinite set of constraints on the boundary OPE coefficients following from locality. In Section \ref{sec:flat}, we show that AdS form factors become flat space form factors in a certain limit of large AdS radius and large scaling dimensions; we derive a phase shift formula for the flat space form factors in physical kinematics in terms of the CFT data. Section \ref{Disc} is a discussion of our results and future directions.

    Appendices \ref{inti} and \ref{supp} contain supplementary formulae omitted for brevity from Sections \ref{localS} and \ref{P4} respectively. Appendix \ref{app:flat} contains certain intermediate formulae for the flat space limit in Section \ref{sec:flat} with general values of the parameters.

    \section{AdS locality and the BOE \la{SAdS}}
    \subsection{Kinematics\label{AdS}}
	We are interested in studying
 quantum fields in AdS space and their dual boundary description. It will sometimes be useful to work in the Poincar\'e patch of AdS (which misses only a point in Euclidean signature),
    \ba
    \ud s^2=R^2 \, \frac{\ud u^2+\ud x_\mu \ud x^\mu}{u^2}\ ,\qquad  \mu=1,\ldots d \ , \la{Poinc}
    \ea
    or in global coordinates,
    \ba
    \ud s^2=\ud r^2+R^2 \sinh^2\left(\frac{r}R\right) \ud \Omega_d^2 \ . \la{glob}
    \ea
    It will 
    be helpful
    to use the embedding space formalism (see e.g. \cite{Penedones:2010ue}), where we think of Euclidean AdS$_{d+1}$ as a hyperboloid in $d+2$ Lorentzian flat space, and its boundary as the (forward) projective null cone:
    \ba
    X^M&\in \mbox{AdS}& &\Leftrightarrow \quad X^2\equiv \eta^{MN} X_M X_N=-R^2&\,,\quad X^0&>0\ , \\
    P^M&\in \partial\mbox{AdS}& &\Leftrightarrow \quad P^2\equiv \eta^{MN} P_M P_N=0&\,,\quad P^0&>0\ ,
    \ea
    with $\eta^{MN}$ the Minkowski metric, $M=0,\ldots,d+1$ and the identification $P^M \sim \lambda P^M$ for real positive $\lambda$. 
In this language, conformal transformations (viz.\ AdS isometries) are simply $d+2$ dimensional Lorentz transformations, and hence conformal invariance of various expressions becomes trivial to check.

    In embedding space, we can choose different `gauges', or parametrizations of AdS, by introducing a fixed future-pointing vector $I$ (which could be null), and setting
\ba
-X\cdot I>0\qquad -P\cdot I=1 \ .
\ea
    For example, the Poincar\'e patch \rf{Poinc} of AdS as well as its global description \rf{glob} (in terms of the Poincar\'e ball) can be obtained respectively by setting
\ba
I^M&=(1,-1,\underbrace{0,\ldots,0}_{d})& \qquad \Rightarrow&\quad \left\{ \begin{array}{l}
X^M=\frac{R}{u}\left(\frac {1+u^2+x^2}2,\frac{1-u^2-x^2}2,x^\mu\right)\,, \vspace{0.3cm}\\
P^M=\left(\frac{1+x^2}{2},\frac{1-x^2}2,x^\mu\right)
\end{array}\right. \\
\vspace{0.2cm} & \\
I^M&=(1,\underbrace{0,\ldots,0}_{d+1})& \qquad \Rightarrow&\quad \left\{ \begin{array}{l}
X^M=R\left(\cosh\left(\frac{r}R\right),\sinh\left(\frac{r}R\right)n_X^a\right)\,,\vspace{0.4 cm} \\
P^M=\left(1,n_P^a\right)
\end{array}\right.
\ea
with $n^a$ unit vectors in $d+1$ dimensions. In conformally invariant observables the dependence on $I$ always drops out, so we will be able to simultaneously describe both cases.

We will actually need very little of the above formalism in our work, although it will simplify several formulae. The reader who is at a loss may just note the following particularly pertinent formulae in the Poincar\'e patch:
\ba
-2P_1\cdot P_2=(x_1-x_2)^2\,, \qquad -2P_1\cdot X=\frac{(x_1-x)^2+u^2}{u} \ .
\ea

We are interested in observables involving both bulk fields $\Fi$ and boundary primary operators $\mathcal O$, written in embedding space as\footnote{More precisely, boundary operators are given by $(-P\cdot I)^{\Delta_\cO} \mathcal O(P)$.}
\ba
\Psi=\Fi(X)\,, \qquad \mathcal O=\mathcal O(P) \,, \qquad 
\cO(\lambda P)= \lambda^{-\Delta_{\cO}} \cO(P) \ .
\ea 
Note that $\Fi$ may denote composite bulk operators and, since our construction will be non-perturbative, we make no \textit{a priori} distinction between elementary and composite operators. However, when $\Fi$ denotes a free field in the bulk, we will denote it $\Phi$, with the corresponding boundary operator denoted $\phi$. 
Finally, we will focus here on locality constraints involving bulk \textit{scalar} fields; however, the same logic applies more generally to spinning fields, which will be considered in the near future \cite{part3}.

\subsection{The boundary operator expansion}
One of the most important properties of a conformal field theory is the state-operator correspondence, which associates eigenstates $|\Delta\rangle$ of the CFT Hamiltonian on the cylinder to local operators. This mapping extends to QFTs in AdS \cite{QFTinAdS}, with the argument going as follows. Time evolution in global AdS (seen as a solid cylinder) maps to rescalings in the Poincar\'e patch.\footnote{
The solid cylinder is obtained from the Poincar\'e patch by setting $x^\mu=e^\tau \tanh(r) n^\mu$, $u=e^\tau \mbox{sech}(r)$, with $n^2=1$, leading to the metric
\ba
\ud s^2=\cosh^2 r\, \ud \tau^2+\ud r^2+\sinh^2 r\, \ud \Omega_{d-1}^2\,.\nonumber
\ea
In particular $\partial_\tau=u \partial_u+x^\mu \partial_\mu$.
}
In particular, a fixed time slice in global coordinates maps to a geodesic surface with constant $u^2+(x-x_0)^2$ for some $x_0$. By evolving states backwards in time, these surfaces become closer and closer to the boundary, where they project onto smaller and smaller boundary spheres. In this way, energy eigenstates in AdS can be associated to localized boundary insertions. In turn the trivial transformation of eigenstates under global time translations tells us that these boundary insertions must have definite transformation properties under dilatations: in particular, they behave as primary or descendant operators. We conclude that the Hilbert spaces of the AdS QFT and the boundary CFT are identified, with the set of primary and descendant boundary CFT states forming a complete basis.
    
\begin{figure}
    \centering
    \includegraphics[width=13cm]{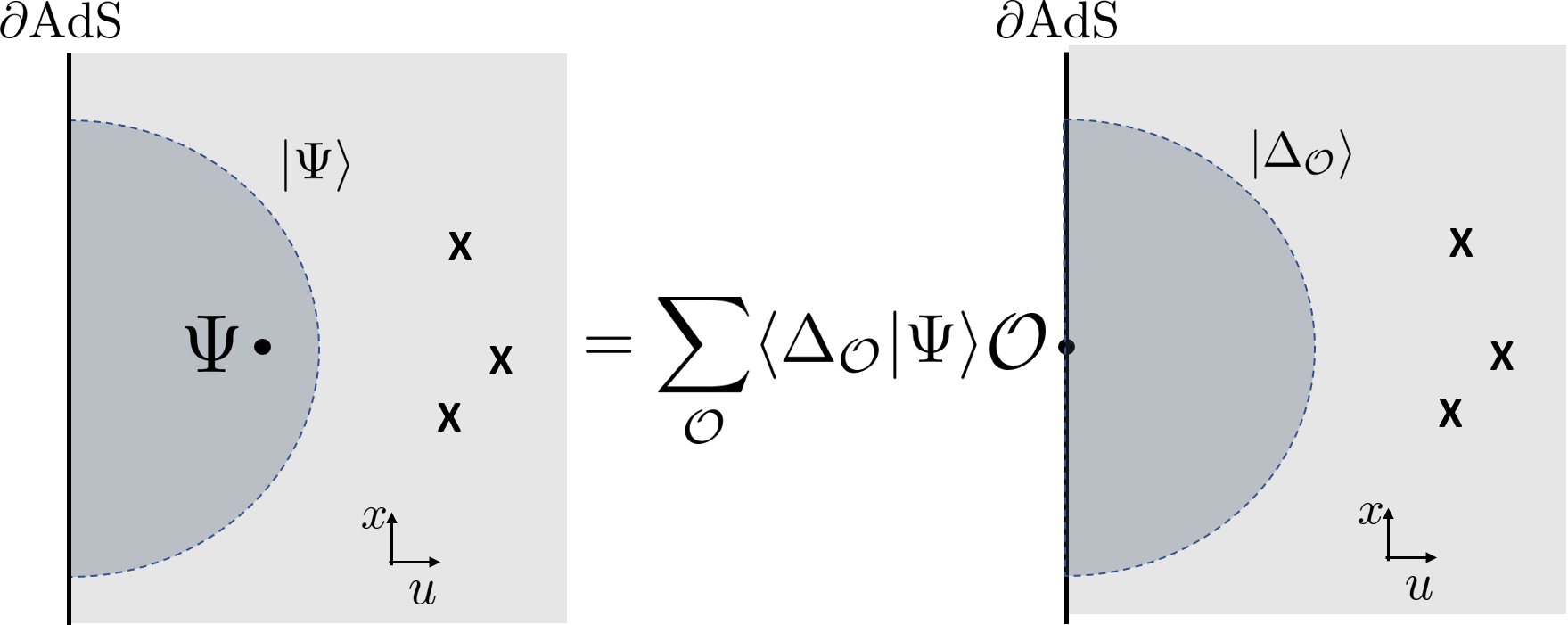}
    \caption{Boundary operator expansion. A bulk insertion creates a state $\Psi$ on the dashed line (a geodesic surface of constant $u^2+x^2$). The state can be expanded in radial quantization, giving us an expression for the operator $\Psi$ in terms of boundary operators $\mathcal O$. This expression holds in correlators with insertions (represented as crosses) outside that surface.
    }
   \label{fig:stateop}
\end{figure}

\begin{figure}
    \centering
    \includegraphics[width=8cm]{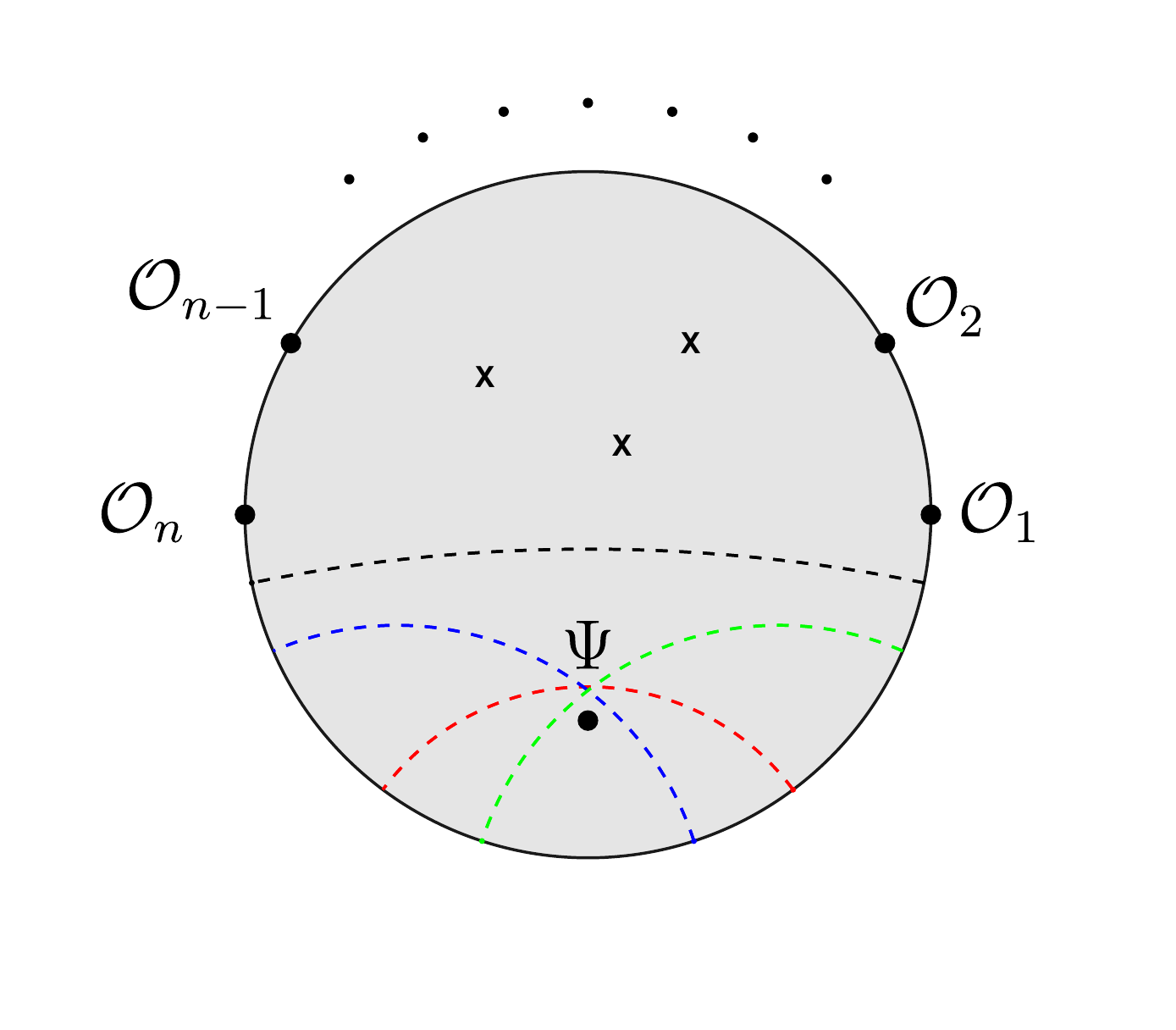}
    \caption{Representation of a correlator involving a bulk field $\Psi$, and other bulk and boundary operators. The BOE holds as long as we can draw a geodesic hypersurface separating $\Psi$ from all other insertions. A few are shown above as dashed lines. Note that to each there corresponds a generically distinct point around which to do radial quantization on the boundary.
    }
   \label{fig:horo}
\end{figure}

The state-operator correspondence guarantees that any state in the bulk may be expressed by acting with operators of definite scaling dimension on the vacuum. In particular, this includes the state obtained by acting with a local bulk operator $\Fi(X)$, so that 
\ba
\langle 0|(\ldots)\Fi(X)|0\rangle=\sum_{\Delta} \mu^\Psi_{\Delta}(X)\, \langle 0|(\ldots)|\Delta\rangle=\sum_{\Delta} \mu^\Psi_{\Delta}(X) \, \langle0|(...)\mathcal O_\Delta(0)|0\rangle \la{BOE} \ ,
\ea
where the dots represent other insertions. The above is known as the {\em boundary operator expansion} (BOE), and is represented diagramatically in figure \ref{fig:stateop}. Note that, since it amounts to decomposing states into an orthonormal basis in a Hilbert space, the BOE converges absolutely. More precisely, this is true as long as we can insert the basis decomposition in the correlator. This means that it must be possible to draw a geodesic hypersurface separating $\Fi(X)$ from all other insertions. Such surfaces are given by those AdS points $X'$ such that $W\cdot X'=0$ for some fixed $d+2$ dimensional spacelike vector $W$. As mentioned above, on the Poincar\'e patch these are surfaces of constant $u^2+(x-x_0)^2$ (see figure \ref{fig:horo}).

The BOE \rf{BOE} leads to the following expressions for bulk fields in terms of boundary primaries (respectively in the Poincar\'e patch and  global coordinates):\footnote{We can also write down a covariant representation
as
\ba
\Fi(X)=\sum_{\Delta} \mu^{\Phi}_{\Delta} \, a^{-\Delta} \widehat C_{\Delta}(b \Box_P) \mathcal O(P) \ , \no
\ea
where we wrote $X^M=a (P^M-b I^M/2)$ and
\ba
\ \Box_P=D_{MN}D^{MN}\,, \quad D_{MN}=P^M\frac{\partial}{\partial P^N}-P_N\frac{\partial}{\partial P^M}  \ .\no
\ea
}
\ba \la{BOE2}
\Fi(u,x)&= \sum_{\Delta} \mu^{\Fi}_{\Delta}\, u^{\Delta} \widehat C_{\Delta}(u^2 \Box_x) \mathcal O_{\Delta}(x)\ ,\\
\Fi(r,n)&=\sum_{\Delta} \mu^{\Fi}_{\Delta}\, [\sinh(r)]^{-\Delta} \widehat C_{\Delta}(t(r) \nabla^2_{S^d}) \mathcal O_{\Delta}(n)
\ea
with $t(r)=2(1-\coth(r))$ and $\nabla^2_{S^d}$ the Laplacian on the $d$ dimensional sphere. 
A couple of comments are in order. The sum $\D$ runs only over scalar primary boundary operators. The differential operator $\widehat C_{\Delta}$ accounts for contributions of descendants. It can be determined by matching the BOE with the two point function of a bulk field and a boundary primary, whose form is in turn fixed by symmetry:
	\ba
	\langle \mathcal O_{\Delta}(P) \Fi(X)\rangle= \frac{\mu_{\Delta}^{\Fi}}{(-2 P\cdot X)^{\Delta}}\qquad \Rightarrow \quad \widehat C_{\Delta}(x)=\sum_{k=0}^\infty \frac{(-1)^m}{\left( \Delta-\frac{d-2}2\right)_m} \left(\frac{x}4\right)^m \ ,
    \ea
    This reasoning also explains why only scalar primary operators can appear in the BOE, as there is no conformally invariant two point function between a bulk scalar and a boundary spinning primary.
% Hence .

%
\subsection{Locality constraints}
An important constraint on correlators of bulk AdS fields is that they must be local. Concretely, this means that singularities in correlation functions may only appear at coincident points in Euclidean signature, or when insertions are null or time-like separated in Lorentzian signature. As we will see, locality implies that BOE coefficients of putative AdS fields must be carefully tuned to avoid unphysical singularities \cite{Kabat:2016zzr}.

\begin{figure}
    \centering
    \begin{tabular}{lr}
    \includegraphics[width=5cm]{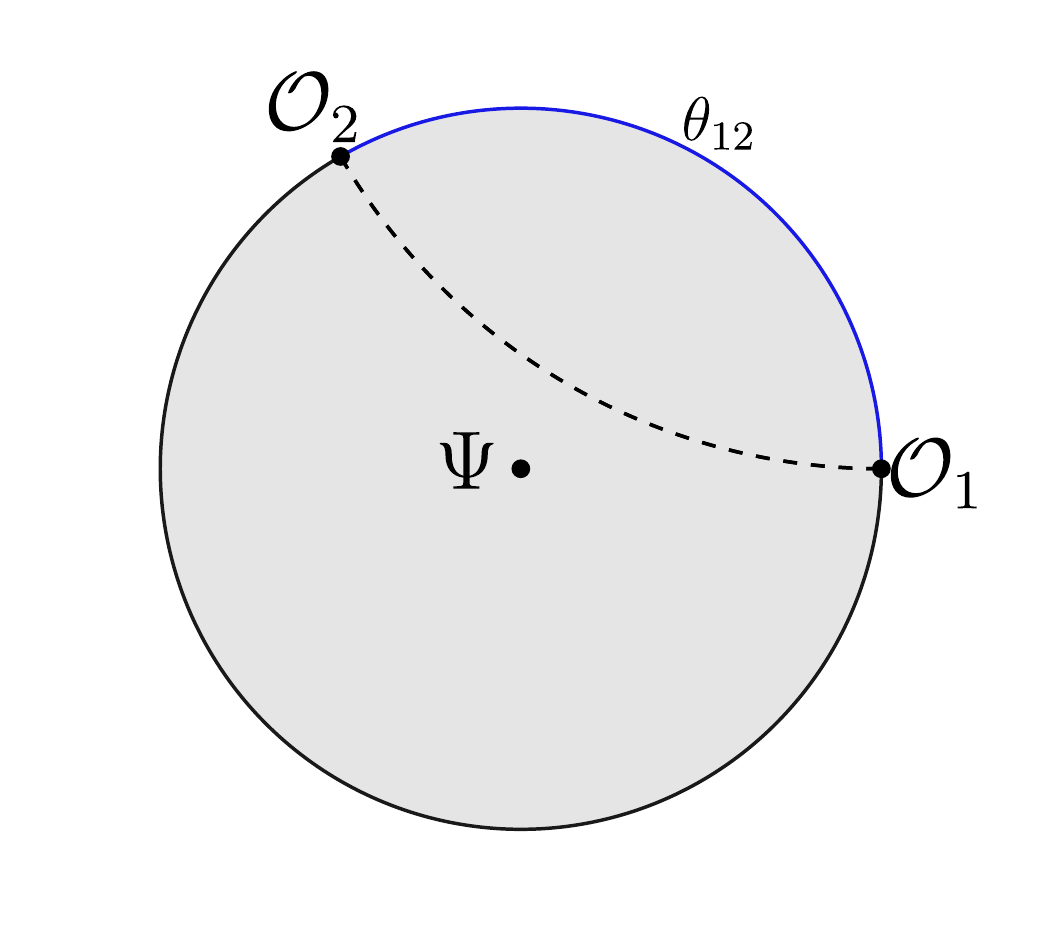}\hspace{1cm}
    &
    \includegraphics[width=9cm]{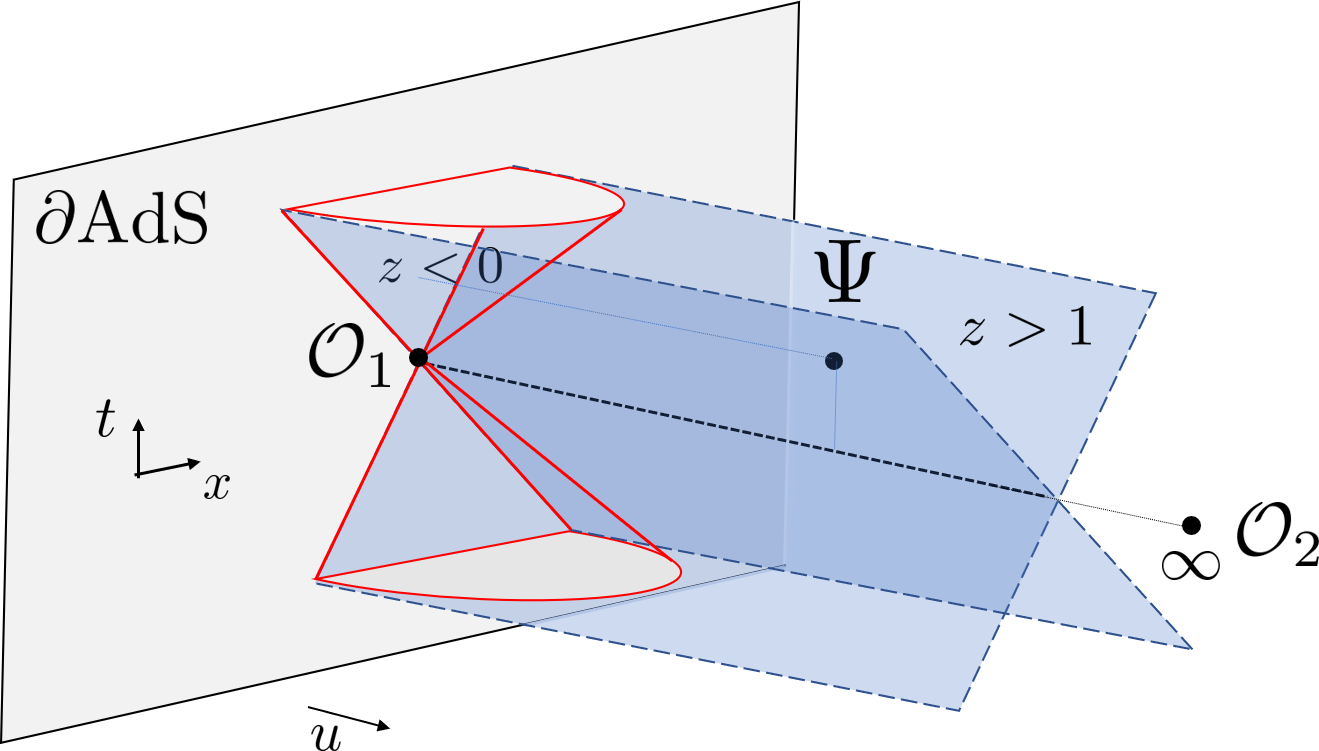}
    \end{tabular}
    \caption{Breakdown of the BOE in a two point AdS form factor. Left: The bulk operator is inserted at $r=0$ in global coordinates and the figure shows an Euclidean AdS$_2$ slice of the full geometry. The two boundary insertions subintend an angle $\theta_{12}$ on the boundary, with $z=\sin^2(\theta_{12}/2)$. The dashed line represents a geodesic connecting these operators. When $\Psi$ lies on this geodesic there can be no hypersurface separating it from $\cO_1$ and $\cO_2$. In the figure this happens when $\theta_{12}=\pi \Leftrightarrow z=1$.  Right: The BOE is also invalid for $z\geq 1$. Such values can be reached in Lorentzian signature, by making the bulk operator timelike separated from the geodesic connecting $\cO_1$ and $\cO_2$ while remaining spacelike separated from both. If the latter condition fails then $z\leq 0$ and the BOE still converges.
    }
    \label{fig:boebreak}
\end{figure}

The simplest correlators subject to non-trivial constraints from bulk locality are mixed 3-point functions $\langle \Psi \, \cO_1 \, \cO_2\rangle$, with one bulk and two boundary insertions. In this case the AdS isometries allow for one conformally invariant cross-ratio $z$, 
\ba
z=-\frac{P_{12}}{2 (P_1\cdot X)( P_2\cdot X)} \qquad \left(=\frac{u^2 (x_1-x_2)^2}{[(x-x_1)^2+u^2][(x-x_2)^2+u^2]}\right) \ .
\ea
To get some intuition for the cross-ratio's meaning, in the Poincar\'e patch, the AdS isometries can be used to fix $u=1,\, $ $x_1^\mu=-x_2^\mu=\rho n^\mu,\, $ $n^2=1$ (see figure \ref{fig:rhovar}). The cross-ratio is then \cite{Hogervorst:2013sma,Bianchi:2022ulu}:
\begin{figure}
    \centering
    \includegraphics[width=5cm]{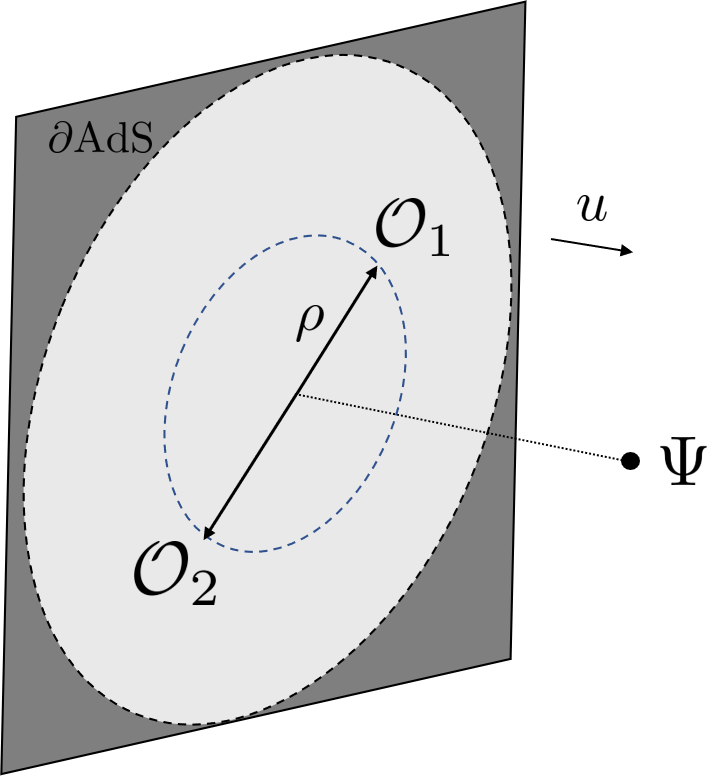}
    \caption{The $\rho$ variable. The operators $\cO_1,\cO_2$ are inserted diametrically opposite at $\rho,-\rho$, and $\Psi=\Psi(x=0,u=1)$. The larger dashed circle has unit radius and represents the boundary projection of the hypersurface $u^2+x^2=1$ passing through $\Psi$.}
    \label{fig:rhovar}
\end{figure}
\ba
z=\frac{4\rho^2}{(1+\rho^2)^2}\,.\label{eq:rho}
\ea
We will use this $\rho$ variable when discussing analytic properties in Section \ref{aff}. Alternatively, focusing on a global AdS$_2$ slice and setting $\Fi$ at its `center' ($r=0$) we get
\ba
z=\sin^2\left(\tfrac{\theta_{12}}2\right)%\frac{1-\cos(\theta_{12})}{2}
\ea
with $\theta_{12}$ the angle between operators on the boundary $S_1$ of the AdS$_2$ slice (see figure \ref{fig:boebreak}). In particular, this makes manifest that, for Euclidean kinematics, the range is $z\in [0,1]$.

A general three point function can be written in the form:
\ba
\langle \Fi(X)  \, \mathcal O_{\D_1}(P_1) \, \mathcal O_{\D_2}(P_2) \rangle  = \frac{1}{(-2P_{12})^{\frac{\D_1+\D_2}2}}\,\left(\frac{P_2\cdot X}{P_1\cdot X}\right)^{\frac{\D_{12}}2}\, F^{\Psi}_{\cO_1\cO_2}(z) \ ,
\ea
where we introduced
\ba
\Delta_{12}\equiv \Delta_1-\Delta_2 \ .
\ea
The function $F(z)$, which we call the `2-point AdS form factor', (or `form factor' for short), will be our main object of study. Using the BOE \rf{BOE2}, it is possible to express the form factor in terms of a \textit{boundary  block expansion}:
\begin{align}
&F^{\Fi}_{\cO_1\cO_2}(z) = \sum_{\cO\in \cO_1\times \cO_2} \mu_\cO^\Fi \, \l^{\cO_1\cO_2}_{\cO} \ G_{\D_{\cO}}^{12} (z) \ , \la{FE}\\
&G_\D^{12}(z) = z^{\frac{\D}2} \,  {}_2{F}_1 (\tfrac{\D+\D_{12}}{2} , \tfrac{\D-\D_{12}}{2}; \D+1-\tfrac{d}{2}; z) \ .\la{blocks}
\end{align}
Each \textit{boundary block} $G_\D^{12}(z)$ captures the contribution of one boundary primary and its descendants to the BOE. Small scaling dimension primaries dominate the BOE when the bulk field approaches the boundary or when the two boundary operators approach each other (i.e.\ in the limit $z\to 0$). 

\begin{figure}
    \centering
    \includegraphics[width=10cm]{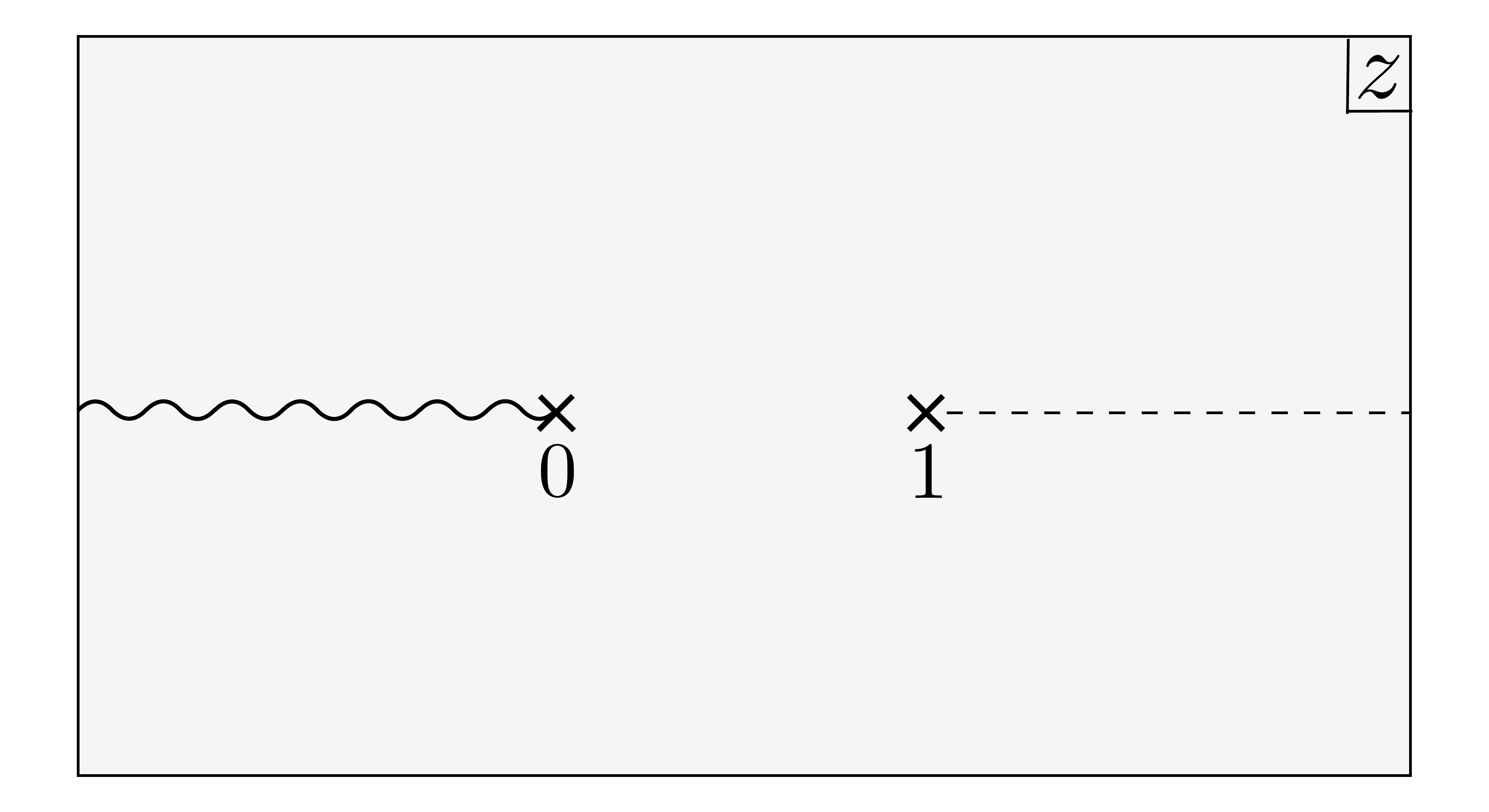}
    \hspace{0.3cm}
    \includegraphics[width=5cm,trim = 0cm 2cm 0cm 0cm]{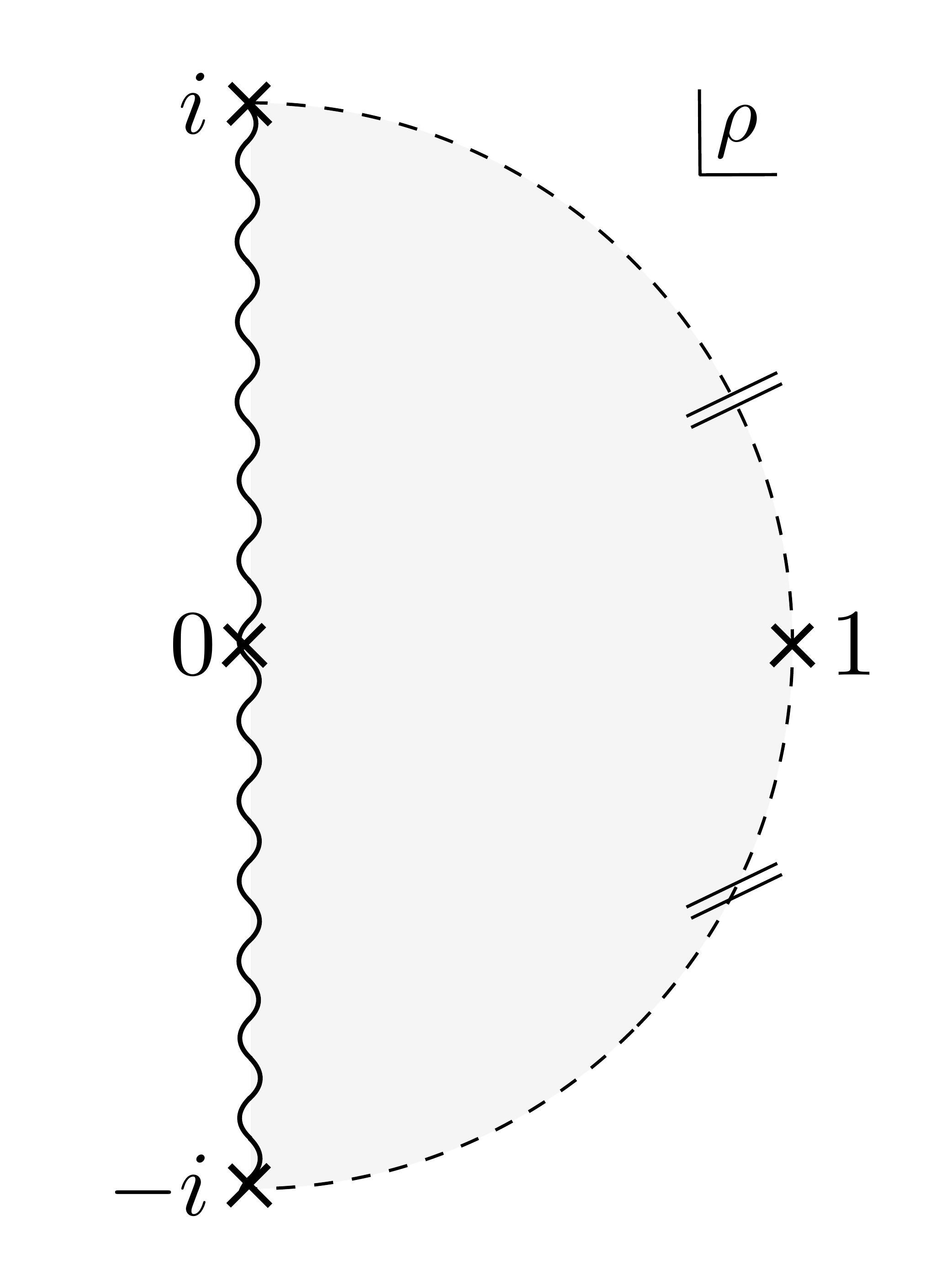}
    \caption{Analytic structure of 2-point AdS form factors $F(z)$ as a function of the complexified cross-ratio. In the $z$ variable, bulk locality only allows a branch cut at $z\leq 0$. In particular, the singularities of the individual boundary blocks at $z\geq 1$ must cancel out in the full form factor. 
    In the $\rho$ variable, the cut plane is mapped to the half disc, with the cut $z\leq 0$ now opened up to cuts running along the imaginary axis; locality forces an identification between the right-hand boundaries above and below $1$: $F(\rho)=F(\rho^*)$ for $|\rho|=1$.}
   \label{fig:branchcuts}
\end{figure}
In Euclidean signature, locality allows a singularity at $z=0$, and in Lorentzian at $z\leq 0$ (see figure \ref{fig:branchcuts}). For all other configurations, operators are spacelike separated, and so must commute. In particular, this implies\footnote{We define the discontinuity $\mathcal I_z F(z)\equiv \lim_{\epsilon\to 0^+} \frac{F(z+i\epsilon)-\bar F(z-i\epsilon)}{2i}$. Note that $F(z)$ must also have no discontinuity for $z\in(0,1)$, but this is manifest from the BOE \rf{blocks}, since the blocks are continuous there.
}
\ba
\mathcal I_z  \, F^{\Fi}_{\cO_1\cO_2}(z) = 0\,, \qquad \mbox{for}\ z\geq 1\,.
\ea
This equation is a non-trivial constraint on the BOE and OPE coefficients, since the boundary blocks {\em do} have a discontinuity at $z\geq 1$, given below in \rf{eq:discG1}. This discontinuity of the blocks is related to the breakdown of convergence of the BOE: when $z=1$ it is impossible to draw a hypersurface separating the bulk field from the two boundary insertions. This extends in Lorentzian signature to the entire region $z\geq 1$ (see figure \ref{fig:boebreak}).
Locality therefore demands that this discontinuity must cancel out in the full BOE \rf{FE}.  

Since the choice of operators $\cO_1$ and $\cO_2$ was arbitrary, the locality constraints must hold for any three-point function. This can only be possible if both the BOE coefficients and the boundary CFT data are carefully tuned. Of course, if this data arises from a QFT in AdS to begin with, such fine tunings are automatic. Here our perspective is that of the inverse problem: starting from given boundary CFT data and attempting to reconstruct local AdS fields (or conversely establishing that no such fields can exist).
We can thus think of bulk locality as the following bootstrap problem:
\vspace{0.2cm}
\begin{flushleft}
{\bf Locality bootstrap:} ({\em Kabat, Lifschytz} \cite{Kabat:2016zzr})\\\vspace{0.3cm} \mbox{{\em Constrain $\Delta_i$, $\lambda^{ij}_{k},$  $\mu_{\Delta_i}^\Fi$ such that, for all $i,j$}:}
\ba
\mathcal I_z \left[\sum_{k} \mu^{\Fi}_{\Delta_k}\, \lambda^{ij}_k \, G_{\Delta_k}^{ij}(z)\right]=0\,, \qquad \mbox{for}\ z\geq 1 \ . \la{lboots}
\ea
\end{flushleft}
\vspace{0.2cm}
Our job will be to turn these constraints into useful equations. The difficulty is that the BOE is not guaranteed to converge for $z\geq 1$, so we may not directly commute the discontinuity above with the infinite sum. In fact, as we show in Section \ref{SR}, we can still extract useful sum rules on the BOE data by smearing the equation above with suitable kernels. Alternatively, and equivalently, we can use the analyticity properties of the form factor to write down a dispersion relation for it, effectively expressing the discontinuity at $z\geq 1$ in terms of its values in other regions of the complex $z$ plane where the BOE does converge. The net result will be a list of sum rules that apply directly to the BOE data.

\subsection{Special cases \la{spec}}

Before we proceed, let us note that there is one possibility for the equations \rf{lboots} to be immediately useful: when the BOE contains only a finite number of blocks. In this case the discontinuity can trivially be commuted with the sum. The blocks' discontinuity is given by%
\ba
\mathcal I_z \, G_\D^{12}(z)&\underset{z\geq 1}{=} \frac{\pi  \Gamma \left(-\frac{d}{2}+\Delta +1\right)}{\Gamma \left(\frac{\Delta +\Delta
   _{12}}2\right) \Gamma \left(\frac{2-d+\Delta -\Delta _{12}}2\right)} \left(\frac{z-1}{z}\right)^{1-\frac d2} z^{-\frac{\Delta_{12}}2}\,J_{\frac{\Delta-2-\Delta_{12}}2}^{(1-\frac d2,\Delta_{12})}(\mbox{$\frac{2-z}z$})\\
   &\underset{z\to 1^+}{\sim} (1-z)^{\frac{2-d}2}\, \frac{\pi  \Gamma \left(1-\frac{d}{2}+\Delta\right)}{\Gamma \left(2-\frac{d}{2}\right) \Gamma \left(\frac{
   \Delta -\Delta _{12}}2\right) \Gamma \left(\frac{
   \Delta +\Delta _{12}}2\right)}
   \label{eq:discG1}
\ea
with $J$ a Jacobi function of the first kind.  Since the Jacobi functions are solutions to a differential equation with generically distinct eigenvalues, they are linearly independent, so that no finite linear combination of them vanishes. There is one exception though: under $\Delta\to d-\Delta$ the particular Jacobi function \rf{eq:discG1} above is left invariant.\footnote{The eigenvalue is $\Delta(\Delta-d)$, as we will see in the next section.} Thus, solutions to locality with a finite number of blocks can occur under only two circumstances: either we tune $\Delta$ so that the prefactor in the equation above is zero; or we combine two blocks with dimensions $\Delta,d-\Delta$ with appropriate relative coefficient. 

The first case corresponds to setting $\Delta=|\Delta_{12}|$, assuming this is compatible with the unitarity bound ($\Delta=0$ or $\Delta\geq (d-2)/2$). A nice example is when the bulk operator is the identity and the form factor becomes the two point-function of boundary operators (therefore $\Delta_{12}=0$). This is clearly local and corresponds to the contribution of the identity block with $\Delta_{12}=\Delta=0$, which has no discontinuity. This means that equations \rf{lboots} cannot constrain its coefficient. 

\sloppy
The second case is where the bulk field satisfies a free bulk equation of motion,~\mbox{$(\nabla_{\mbox{\tiny AdS}}^2-m^2)\Phi=0$}. In this case the BOE may contain at most two operators with dimensions $\Delta, d-\Delta$ with $m^2=\Delta(\Delta-d)$. Using \reef{eq:discG1}, it is easily seen that the locality equations \rf{lboots} are solved by setting:
\ba
\frac{\mu^\Phi_\Delta \lambda^{12}_{\Delta}}{\mu^\Phi_{d-\Delta} \lambda^{12}_{d-\Delta}}=\frac{ \Gamma \left(\frac{d}{2}-\Delta +1\right)   \Gamma\left(\frac{\Delta+\Delta_{12}}2\right)\Gamma\left(\frac{\Delta-\Delta_{12}}2\right)}{\Gamma\left(\Delta-\frac d2+1\right)\Gamma\left(\frac{d-\Delta+\Delta_{12}}2\right)\Gamma\left(\frac{d-\Delta-\Delta_{12}}2\right)}\,.
\ea
This relation degenerates when $\Delta=|\Delta_{12}|$ or $d-\Delta=|\Delta_{12}|$ in which case, as expected, a single operator appears in the BOE. The associated form factors are $\langle \Phi 
\, \phi^n \, \phi^{n+1}\rangle$  with $n$ arbitrary, $\Phi$ a free bulk field and $\phi$ its dual boundary operator.

To conclude, note that by considering two or more distinct pairs of boundary operators, the above implies relations that only involve the boundary CFT data. Such relations were first derived in \cite{Paulos:2015jfa} for the long range Ising model, which is described by a bulk free field with a boundary interaction. They have since been generalized to arbitrary spins and used in bootstrap applications in \cite{Behan:2018hfx,Behan:2020nsf,Behan:2021tcn}. The results in the present work allow for a generalization of this logic to arbitrary BOEs, something which will be briefly explored in Section \ref{loccft}.

\section{The local block expansion \la{Sloc}}
\subsection{Analyticity properties of form factors \la{aff}}
In this section, we will study the analyticity properties of form factors.  We begin by decluttering the notation, writing a general form factor as:
\ba
F(z)=\sum_{\Delta\geq 0} c_{\Delta} \, G^{12}_{\Delta}(z)\ , \qquad \qquad c_\D = \mu_\D^{\Psi} \l_\D^{12}\,. \la{BOE22}
\ea
Concretely, we have dropped the explicit dependence on $\Fi, \mathcal O_1,\mathcal O_2$. We will take all operators to be real, implying reality of the coefficients $c_{\Delta}$. Note, however, that these coefficients are not sign-definite. The sum over states ranges over the primary operators in the boundary CFT but, since we want to remain agnostic about the precise CFT under consideration,  we allow all possible values of $\Delta$ that are consistent with unitarity.\footnote{The scaling dimensons allowed by unitarity are $\Delta=0$ and $\Delta\geq \frac{d-2}2$ for $d\geq 2$ and just $\Delta\geq 0$ for $d\leq 2$.}

Before the form factor itself, let us discuss the properties of boundary blocks \rf{blocks}. The easiest way to compute the blocks is to notice that they satisfy a Casimir equation. This follows from the fact that scalar operators satisfy by construction
\bea
\nabla^2_{\mbox{\tiny AdS}} \Psi(X)=[C_{(2)},\Psi(X)] \ ,
\eea
where $C_{(2)}$ is the CFT quadratic Casimir operator.\footnote{Note that the AdS Laplacian can be written in embedding formalism as $\mathcal M^{MN}\mathcal M_{MN}$, with $\mathcal M_{MN}=X_{[M} \partial_{N]}$.}
A boundary block captures the contribution to the bulk field from a single boundary primary, corresponding to a piece of the bulk field that satisfies a free AdS wave equation. In our form factor context, the action of the AdS wave operator $\nabla^2_{\mbox{\tiny AdS}}-\Delta(\Delta-d)$ yields:
\ba
\left[C^{12}_z-\lambda_{\Delta}\right]G_{\Delta}^{12}(z)
=0
\ea
with 
\be
C^{12}_z := 4(1-z)^{1-d/2}z^{1+d/2} \partial_z [ (1-z)^{d/2}z^{1-d/2} \partial_z ] +  \D_{12}^2 z
 \ , \qquad \lambda_\D = \D(\D-d) \ .\label{eq:cas}
\ee
The blocks correspond to the solutions of this equation satisfying the asymptotics\footnote{The asymptotics may suggest that the $d=1$ blocks are analytic for $z>1$, but this is not the case, since for $d<2$ blocks still contain a (subleading) term $(1-z)^{\frac{(2-d)}2}$ in their expansion.}
\ba
G^{12}_{\Delta}(z)&\underset{z\to 0^+}{=} z^{\frac{\Delta}2}+\ldots\, ,\qquad |G^{12}_{\Delta}(z)|\underset{z\to\infty}=O(z^{\frac{|\D_{12}|}2}) \quad \left(\underset{\D_1=\D_2}=O(\log(z))\right) \ ,\\
G^{12}_{\Delta}(z)&\underset{z\to 1^-}=\left\{
\begin{array}{ll}
O(1) &, \quad  d=1\\
O(\log(1-z)) &, \quad  d=2\\
O((1-z)^\frac{2-d}2) &, \quad  d>2 \ .
\end{array}
\right. 
\ea
The blocks are analytic functions except for possible singularities for $z\leq 0$ and $z\geq 1$. A proof of this follows from their representation in terms of the BOE, as will be shown below. It will be useful for us to note that blocks satisfy the identity
\ba
G_{\Delta}^{12}(z)=e^{i\frac{\pi}2\Delta}(1-z)^{-\frac{|\D_{12}|}2} \tilde G_{\Delta}^{12}(\mbox{$\frac{z}{z-1}$})\,,\qquad \mbox{Im}~z>0  \ , \la{eq:idblock}
\ea
where
\ba
\tilde G_{\Delta}^{12}(z):= z^{\frac{\D}2} \,  {}_2{F}_1 (\tfrac{\D+|\D_{12}|}{2} , \tfrac{\D+2-d+|\D_{12}|}{2}; \D+1-\tfrac{d}{2}; z) \ .
\ea
% Furthermore, we point out that $\tilde G_{\Delta}^{12}(z)$ is positive for $0<z<1$ and $\Delta$ satisfying the unitarity bound. This will be important below.

Let us now understand the analyticity properties of the form factor. We begin by writing it in the form of an expansion:
\begin{align}
F(z)&=\sum_{\Delta}\sum_{n=0}^\infty z^{\frac{\Delta+2n}2} \,  c_{\Delta,n}^{12} \la{e1}\\
&=\sum_{\Delta}\sum_{n=0}^\infty \rho^{\Delta+2n} \, \tilde c_{\Delta,n}^{12} \ , \quad \qquad \rho^2(z):=\frac{1-\sqrt{1-z}} {1+\sqrt{1-z}} \  \la{e2}
\end{align}
for some coefficients $c_n$ and $\tilde c_n$. It turns out the latter expansion \rf{e2} has a simple Hilbert space interpretation. To see this, let us first set $d=1$ for clarity of presentation. Inserting a complete set of states in the form factor gives\footnote{For the derivation note that: 
\ba
\nonumber
\langle 0|\Psi(1,0) P^{m}|\Delta\rangle&=0\qquad \mbox{for odd}\ m\\
\langle \Delta| K^{2n} \cO_1(\rho)\cO_2(-\rho)|0\rangle&=\rho^{\Delta+2n-\Delta_1-\Delta_2}\langle \Delta|K^{2n} \cO_1(1)\cO_2(-1)|0\rangle.
\ea}
\ba
F(z)&=(2\rho)^{\Delta_1+\Delta_2}\langle 0| \Psi(u=1,x=0) \cO_1(\rho ) \cO_2(-\rho)|0\rangle\\
&=(2\rho)^{\Delta_1+\Delta_2}\langle 0| \Psi(1,0)\left(\sum_{\Delta,m} \frac{P^m |\Delta\rangle\langle \Delta|K^m}{\langle \Delta|K^m P^m|\Delta\rangle}\right) \cO_1(\rho) \cO_2(-\rho)|0\rangle\\
\Rightarrow \tilde c_{\Delta,n}^{12}&=\frac{2^{\Delta_1+\Delta_2}\langle 0| \Psi(1,0) P^{2n} |\Delta\rangle \langle \Delta|K^{2n}\cO_1(1) \cO_2(-1)|0\rangle}{\langle \Delta|K^{2n} P^{2n}|\Delta\rangle}\,.
\ea
Hence the $\rho$ expansion directly corresponds to the BOE expansion of the form factor. The above shows this is an orthonormal basis decomposition of an overlap between two states in a Hilbert space, and as such it converges absolutely. Thus, convergence for $\rho<1$ (which was necessary for inserting the basis decomposition) is promoted to $|\rho|<1$. Translating back to the $z$ variable, this implies that $F$ is analytic on (a multi-sheeted cover of) the complex $z$ plane with a branch cut running along $z\leq0$ (see figure \ref{fig:branchcuts}). From the BOE alone, we cannot make any statements about $|\rho|=1$ (i.e.\ $z\geq 1$): analyticity there is precisely the property of locality that we wish to study.
\unskip\footnote{As a side note, since the mapping from $\rho$ to $z$ is analytic for $|z|<1$, it follows that the $z$ expansion of the form factor also converges absolutely in that domain. It is likely that this expansion also has a simple Hilbert space interpretation --- there is one for an expansion in $z/(1-z)$ --- but we were unable to find it.}

\begin{figure}
    \centering
    \includegraphics[width=7cm]{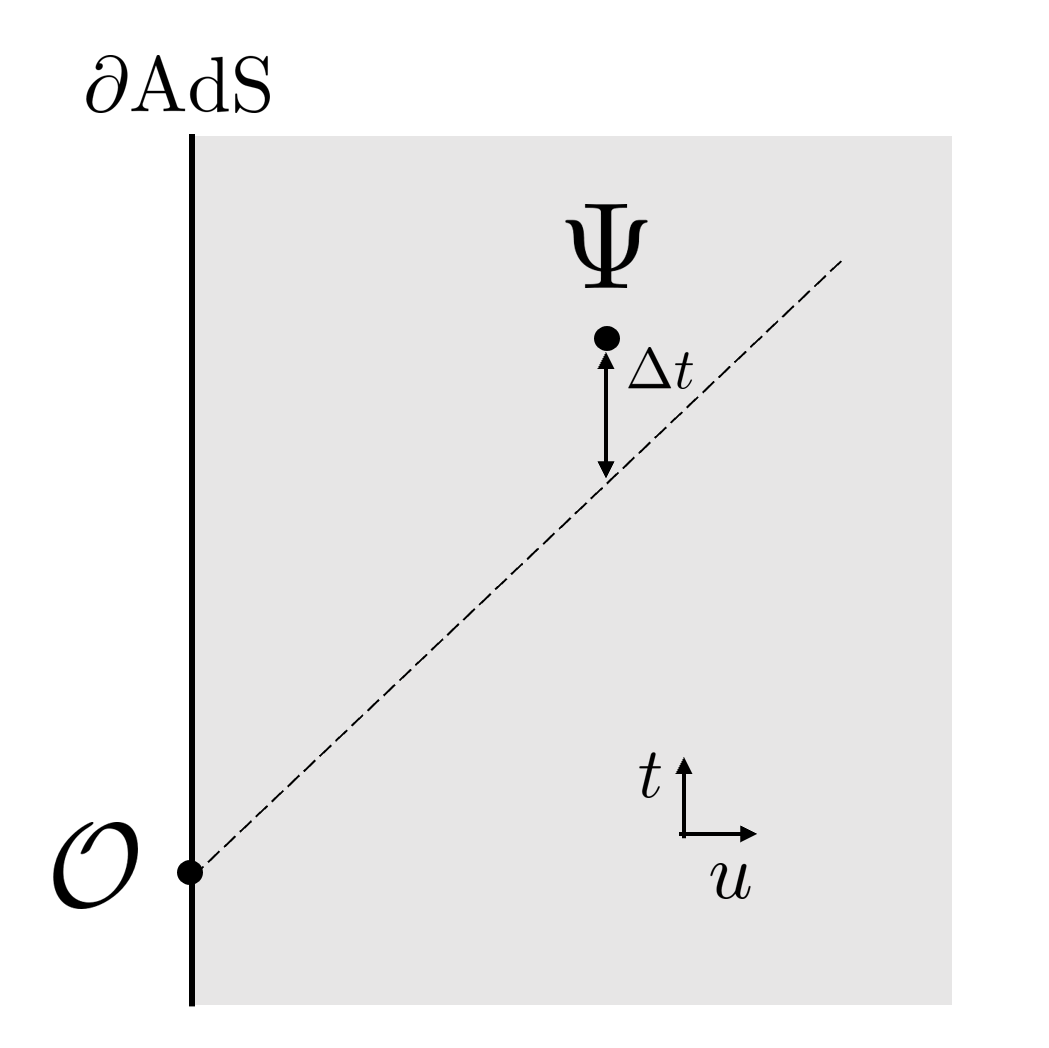}
    \caption{Bulk operator approaching the lightcone of a boundary insertion. In this configuration the crossratio $z\sim -\frac{u}{2 \Delta t}<0$. Excitations traveling from boundary to bulk have energies $E\propto e^{\beta}$ with $\beta\gg 1$ the boost, and relativistic kinematics gives $e^{-2\beta}\propto \Delta t$, so that~$z\sim E^2$.
    }
   \label{fig:regge}
\end{figure}

Finally, it will be important below to constrain the behaviour of the form factor in the limit $z\to \infty$, which corresponds to the bulk field approaching the lightcone of a boundary insertion. Physically we can think of this limit as dominated by the exchange of highly energetic particles between the bulk insertion and the boundary operator. As the bulk field approaches the lightcone, the energy of the emmitted particles goes as $E\sim 1/(\Delta t)^{\frac 12}\sim \sqrt{z}$, as described in figure \ref{fig:regge}.
Since large $z$ corresponds to high energies, in this limit we are sensitive to the flat space form factor of the bulk field. In Section~\ref{sec:flat}, where we study a related limit, we will see that this heuristic argument becomes precise. 

This reasoning motivates the important assumption that form factors should be polynomially bounded as $z\to \infty$. The exponent in the polynomial bound will \textit{a priori} depend on the specific insertions in the form factor, and it should correlate with the expected large momentum limit of the form factor in flat space. We can argue for this more precisely as follows. Given a local form factor $F^{\Psi}$ of an operator $\Psi$, bounded by a power  $\sim z^{\alpha_\Psi}$, it is clear that the form factor of $\nabla^2_{\mbox{\tiny AdS}} \Psi$ will also be local and satisfy
\ba
F^{\nabla^2 \Psi}(z)=C^{12}_z F^{\Psi}(z)\underset{z\to\infty}{=}O(z^{\alpha_\Psi+1})\,,
\ea
i.e.\ with a harder asymptotic. Incidentally, note this is consistent with our identification $z\sim E^2$. Since the action of the Casimir operator maps boundary blocks to themselves, it follows that the states appearing in the BOE of $F^{\nabla^2 \Psi}$ are exactly the same as those appearing in that of $F^\Psi$. As such, in order to characterize a form factor it is not sufficient to specify that it is local and which operators appear in its BOE, but rather we will also have to specify its behaviour for large~$z$.

For our purposes, it will be convenient to impose polynomial boundedness in a specific way. Firstly, from a form factor $F$ we define the related function (cf. \rf{e2}):
\ba
\tilde F(z):=\sum_{\Delta}\sum_{n=0}^\infty \left|c_{\Delta,n}^{12}\right|\rho^{\Delta+2n}\,,\qquad \quad \rho=\rho(z)\,. \la{FTD}
\ea
This is a positive function for $z\in (0,1)$ and analytic for $|\rho|<1$ thanks to absolute convergence of the BOE. We now assume that $\tilde F(z)$ is polynomially bounded as $z\to 1$, which implies a bound on $F$ itself:\footnote{Note that the singularity structure of blocks implies:\ 
$\aF\geq \mbox{max}\left\{\frac{|\Delta_{12}|}2, \frac{d-2}2\right\}$.
}
\ba
(1-z)^{\aF}\tilde F(z)\underset{z\to 1}=O(1)\qquad \Rightarrow \qquad |F(z)|\underset{|z|\to \infty}= O(|z|^{\aF})\ . \la{bas}
\ea
This implication follows from 
\ba
|F(z)|\leq \tilde F(z^+(z))\,, \qquad \mbox{with}\quad z^+(z)\equiv\frac{4 |\rho^2(z)|}{(1+|\rho^2(z)|)^2}\,,
\ea
together with $z^+(z)\underset{z\to \infty}{=} 1-O(1/|z|)$.

To summarize, in this section we have argued that local form factors are analytic functions in the $z$ plane with a branch point at $z=0$. We have also assumed that they are polynomially bounded with an exponent, which is in principle arbitrary and dependent on the specific bulk field under consideration.

\subsection{Dispersion relation and local blocks \la{Sdis}}
In this section, we will use the analyticity properties described above to write down a dispersion formula. Combining this formula with the BOE will lead to general constraints on local form factors.

To derive this formula, we begin by setting:
\be \la{GFFf}
\overline{F}(z)\equiv F(z)-\sum_{0\leq \Delta\leq 2\af} c_{\Delta} \mathcal L^{12,\af}_{\Delta}(z)
\ee
for some $\af$ to be specified below. 
Here the $c_{\Delta}$ are be the same coefficients appearing in the BOE \rf{FE} of $F(z)$ and we have introduced functions $\mathcal L^{12,\af}_{\Delta}$ that we will call  {\em local blocks}. We assume that they have the same analyticity and boundedness properties as $F(z)$ --- i.e. they are local form factors themselves --- but satisfy in addition
\ba
\mathcal L^{12,\af}_{\Delta}(z)\underset{z\to 0^+}=G^{12}_{\Delta}(z)+O(z^{\af+\epsilon}) \ , \qquad \epsilon>0\ . 
\ea
The reason for these definitions is that $\overline F(z)$ also has the properties of a form factor, but is now suppressed near $z=0$ as:
\ba
\overline{F}(z)\underset{z\to 0+}=O(z^{\af+\epsilon})\ .
\ea
The definition of local blocks will become apparent in the course of the derivation. Their role so far is to provide subtractions that will allow us to write our desired dispersion relation. We begin with the Cauchy formula:
\ba
\overline F(w)=\oint \frac{\ud z}{2\pi i} \left(\frac wz\right)^{\af+1}\frac{\overline F(z)}{z-w}
\ea
where the contour encircles the pole at $z=w$. As usual, we want to deform the contour to pick up the discontinuities of the integrand.  In order to drop contributions at infinity we must constrain:
\ba
\af>\aF-1 \ .\la{constr}
\ea
Deforming the contour, we pick up only contributions from discontinuities at $z<0$, since $\overline F$ is assumed to behave as a local form factor even after the subtractions. This gives:
\ba
\overline F(w) =\int_{-\infty}^0 \frac{\ud z}{\pi}\,\frac {w^{\af+1}}{z(z-w)}\mathcal I_z\left[z^{-\af} \overline F(z)\right] \ . \label{eq:dispFo}
\ea
We see that subtractions were necessary so as to obtain a finite integral in the region of small negative $z$. To get a dispersion relation for $F(z)$ we still need to define the local blocks $\mathcal L^{12,\af}_{\Delta}$. To do this, let us first set:
\ba
\mathcal L_{\Delta}^{12,\af}(w)=\int_{-\infty}^0 \frac{\ud z}{\pi}\,\frac {w^{\af+1}}{z(z-w)}\mathcal I_z\left[z^{-\af} G_{\Delta}^{12}(z)\right] \,, \qquad \mbox{for}\ \  \Delta> 2 \af\,. \label{eq:replocal1}
\ea
To obtain the analytic continuation for other values of $\Delta$ (required to define the subtractions), we deform the contour to get
\ba
\mathcal L^{12,\af}_{\Delta}(w)=G_{\Delta}^{12}(w)-\int_{1}^{\infty} \frac{\ud z}{\pi}\,\left(\frac{w}{z}\right)^{\af+1} \frac{\mathcal I_z G_{\Delta}^{12}(z)}{z-w} \,, \label{eq:replocal2}
\ea
which is now valid for any $\Delta$. This expression makes manifest that the local block does indeed have all the analyticity and boundedness properties expected of a form factor. In particular it also implies the crucial property:
\ba
\mathcal I_z\left[ z^{-\af}\mathcal L_{\Delta}^{12}(z)\right]=\mathcal I_z\left[z^{-\af} G_{\Delta}^{12}(z)\right]\,, \qquad z<0\,.
\ea
This allows us to unpack \reef{eq:dispFo} to finally obtain the dispersion formula:
\begin{multline}
F(w)=\sum_{0\leq \Delta\leq 2 \af} c_{\Delta}\mathcal L_{\Delta}^{12,\af}(w)\\+\int_{-\infty}^0 \frac{\ud z}{\pi}\,\frac {w^{\af+1}}{z(z-w)}\mathcal I_z\left[z^{-\af}\left(F(z)-\sum_{0\leq \Delta\leq 2\af} c_{\Delta} G_{\Delta}^{12}(z)\right)\right] \,.\label{eq:dispF}
\end{multline}
A different way of thinking about this formula is more useful: let us plug the BOE \rf{BOE22} into both sides of this equation and, crucially, assume that the sum over states commutes with the dispersive integral. 
In Section \ref{sec:swapping} we will show that polynomial boundedness of $F$ is sufficient to justify this `swapping' property, independently of whether $F$ is actually local. Then we can state:
\ba
\boxed{
\mathcal I_z F(z)\underset{z\geq 1}=0\qquad \Leftrightarrow \qquad 
\sum_{\Delta\geq 0} c_{\Delta} G_{\Delta}^{12}(z)=\sum_{\Delta\geq 0} c_{\Delta} \mathcal L_{\Delta}^{12,\af}(z) \label{eq:localbs}
}
\ea
To prove this, first note that, thanks to  swapping and equations \eqref{eq:replocal1}, \eqref{eq:replocal2}, the equation on the right-hand side is the same as \reef{eq:dispF}. Hence, showing the direct implication ($\Rightarrow$) amounts to deriving \rf{eq:dispF}, which we just did, while the reverse implication ($\Leftarrow$) follows trivially by computing the discontinuity in \rf{eq:dispF}, or alternatively running the derivation of that equation backwards.

Let us now discuss the meaning of \rf{eq:localbs}. It tells us that a form factor is local if and only if it admits admits two expansions: one in terms of boundary blocks $G_\D^{12}$, which manifests the BOE but not locality; and another, in terms of local blocks $\mathcal L_\D^{12,\tilde \a}$, which manifests locality but not the BOE. We note also that the definition of the blocks implies
\ba \la{lbe}
\mathcal L_{\Delta}^{12,\tilde \alpha}(w) =G_{\Delta}^{12}(w)-\sum_{n=1}^\infty \theta_n^{12,\af}(\Delta)\, G^{12}_{2\af+2n}(w) \ ,
\ea
where the coefficients $\theta_n^{12,\af}(\Delta)$ will be found in Section \ref{WA}:
\begin{align}
\theta_n^{12,\tilde \a} ( \D) 
&=\frac{4(-1)^n}{(n-1)!}\frac{ 1}{(\D-2\tilde \a-2n)(\D+2\tilde \a+2n-d)} \la{thr}\\
&\qquad \qquad   \times \frac{\G(\tilde \a+n+ \tfrac{\D_{12}}{2})\G(\tilde \a+n- \tfrac{\D_{12}}{2})\G(\D+1-\tfrac{d}{2})\G(2 \tilde \a +n+1-\tfrac{d}{2})}{\G(\tfrac{\D}{2}+\tfrac{\D_{12}}{2})\G(\tfrac{\D}{2}-\tfrac{\D_{12}}{2})\G(\tilde \a-\tfrac{\D}{2}+1)\G(\tilde \a +\tfrac{\D}{2}+1-\tfrac{d}{2})\G(2 \tilde \a +2n-\tfrac{d}{2})} \ . \no
\end{align}
This suggests an alternative way of formulating the equality of the expansions in \rf{eq:localbs}:
\ba
\boxed{
\mathcal I_z F(z)\underset{z\geq 1}=0 \qquad \Leftrightarrow \qquad \sum_{\Delta} c_{\Delta} \, \theta_n^{12,\af}(\Delta)=0\quad \quad \mbox{for all}\quad n\geq 1 \label{eq:sumrules}
}
\ea
thus rephrasing locality as a discrete set of sum rules, as desired.  Deriving these sum rules requires commuting two infinite series, so a more careful treatment is required: we will properly establish the validity of \rf{eq:sumrules} in the next subsection.

These results are strongly reminiscent of the Polyakov bootstrap \cite{Gopakumar:2016cpb,Gopakumar:2018xqi,Gopakumar:2016wkt,Gopakumar:2021dvg}. The dispersion relation we have found above is the analog of a similar dispersion formula for CFT correlators involving the double discontinuity \cite{Carmi:2019cub,Paulos:2020zxx}. In our case, there is a single discontinuity, but this is still useful.\footnote{See \cite{Bissi:2019kkx} for related work in the context of the CFT four-point function.}
To see this, first note
\ba
\mathcal I_z \left[z^{-\af} G_{\Delta}^{12}(z)\right]\propto \sin\left[\frac \pi2 (\Delta-2\af)\right]\ , \qquad z<0\ ,
\ea
which in turn implies
\ba
\mathcal L_{\Delta}^{12,\af}(z)\bigg|_{\Delta=2\af+2n}=0 \qquad \mbox{for}\quad n\geq 1 \la{vl} \ .
\ea
It follows that by tuning $\af$ we can eliminate contributions from towers of states in the local block representation of the form factor, analogous to the decoupling of double-trace operators in the Polyakov bootstrap. This is especially useful in perturbative computations of form factors, as we will see explicitly in Section \ref{WA}.

\section{Sum rules from functionals \la{SR}}
The basic issue with the constraints implicit in \reef{lboots} is 
the lack of convergence of the BOE along the $z\geq 1$ cut, preventing us from commuting the BOE sum with the discontinuity. In this section, we will show that is possible to get around this by first smearing the sum with suitable functional kernels. The story is similar to what has been done for CFT correlators in \cite{Mazac:2018ycv}. 
\subsection{Conditions on functional kernels}
\la{sec:swapping}
We begin by defining the functional action:
\ba
\omega_f[F]\equiv \int_\Gamma \frac{\ud z}{2\pi i} \, f(z)\,  F(z)\,, \qquad \Gamma=\{z: z=\frac 12+i t,\quad t\in \mathbb R\} \label{eq:basicfuncac}\,.
\ea
We will mostly consider kernels $f(z)$ that are analytic in $\mathbb C\backslash (-\infty,0]$ and with suitable power boundedness at infinity:
\ba
|f(z)|\underset{|z|\to \infty}{=}O(|z|^{-2-\af})\,. \label{eq:asympf}
\ea
We also have in mind their action on functions $F(z)$ with suitable analytic properties. In particular,
we should be able to perform a contour deformation to obtain
\ba
\omega_f[F]=\int_{1}^\infty \frac{\ud z}{\pi}\, f(z) \, \mathcal I_z F(z)
\ea
so that 
\ba
 \omega_f[F]=0 \quad \mbox{if}\ \ F\ \mbox{is local} \ .
\ea
To obtain a useful result, we would like to show that the integral above may be commuted with the BOE of $F$ \cite{Qiao:2017lkv}. That is, we want the following `swapping' property:
\ba
\omega_f[F]=\sum_{\Delta} c_{\Delta} \omega_f^{12}(\Delta)\,, \qquad \omega_f^{12}(\Delta):= \omega_f[G_{\Delta}^{12}] \label{eq:swapping} \ .
\ea
This property follows from our assumptions on $f,F$ and, crucially, from the polynomial boundedness properties we introduced in the previous section. To see this, first note that, with the notation of that section, we have
\ba
\left|f(z) \sum_{\Delta\leq \Delta^*} c_{\Delta} G_{\Delta}^{12}(z)\right| \leq |f(z)| \tilde F(z^+(z))\,, \qquad \mbox{for all}\ \Delta^*\,.
\ea
Furthermore we have
\ba
\int_\Gamma \frac{\ud z}{2\pi} |f(z)|\tilde F(z^+(z))<\infty\,, 
\ea
as easily follows from \reef{eq:asympf},\rf{bas} and the constraint on $\af$, which we repeat here:
\ba
\af>\alpha_F-1\ .
\ea
The dominated convergence theorem now allows the exchange of the series and integration, giving
\begin{multline}
\lim_{\Delta*\to \infty}\sum_{\Delta\leq \Delta^*} \int_{\Gamma} \frac{\ud z}{2\pi i} f(z) c_{\Delta} G_{\Delta}^{12}(z)\\
=\int_{\Gamma}\frac{\ud z}{2\pi i}  f(z) \left[\lim_{\Delta*\to \infty} \sum_{\Delta\leq \Delta^*} c_{\Delta} G_{\Delta}^{12}(z)\right]=\int_{\Gamma}\frac{\ud z}{2\pi i} f(z) F(z) \ ,
\end{multline}
which is the same as \reef{eq:swapping}.

\subsection{Bases of functionals}
\subsubsection{Master functionals}
A simple but important family of functionals satisfying the conditions set out above are so-called {\em master functionals}. Denoted by $\Lambda_w^{\af}$, they are associated with kernels:
\ba
\Lambda_w^{\af}:\qquad f_w^{\af}(z):= \left(\frac{w}z\right)^{1+\af}\, \frac{1}{z-w}\,.
\ea
Notice that this is the same kernel that appeared in the dispersion relation \rf{eq:dispF} of the previous section. Actually, the fact that the master functionals are admissible functionals automatically establishes the swapping property that was mentioned there. Indeed, as we now show, we can derive the local block expansion in \rf{eq:localbs} directly from the master functional sum rule. 

The master functionals act as
\ba
\Lambda_w^{\af}[F]:=\int_1^\infty \frac{\ud z}{\pi}\, f_w^{\af}(z) \mathcal I_z F(z)  \ , \la{eq:masterac}
\ea
and, in particular, on boundary blocks as
\ba
\Lambda_w^{\af}[G_{\Delta}^{12}]\equiv \Lambda_w^{12,\af}(\Delta)=G^{12}_{\Delta}(w)-\mathcal L_{\Delta}^{12,\af}(w)\ , 
\ea
by the definition \rf{eq:replocal2} of the local blocks. Hence the corresponding sum rule is:
\ba
0=\sum_{\Delta} c_{\Delta} \, \Lambda_w^{12,\af}(\Delta)=\sum_{\Delta} c_\Delta\left[ G_{\Delta}^{12}(w)-\mathcal L_{\Delta}^{12,\af}(w)\right]\ \  \Leftrightarrow \ \  F(w)=\sum_{\Delta} c_{\Delta} \, \mathcal L_{\Delta}^{12,\af}(w) \ ,
\ea
i.e. we recover the representation \rf{eq:localbs} of $F$ in terms of local blocks. This means we have proved:
\ba
\mathcal I_z F(z)\underset{z\geq 1}=0\quad \Rightarrow \quad F(z)=\sum_{\Delta\geq 0} c_{\Delta} \mathcal L_{\Delta}^{12,\af}(z) \label{eq:localbs2}\ .
\ea
This is simply the ($\Rightarrow$) implication of eq.\ \rf{eq:localbs} above. For completeness, the converse implication ($\Leftarrow$) follows in the language of this section from deforming the contour in \rf{eq:masterac}, as in the derivation of the dispersion formula.

\subsubsection{Sum rules}
Let us now see how to recover the sum rules appearing in \reef{eq:sumrules}.
We begin by expanding the master functional kernel in powers of the auxiliary variable $w$. In this way, we obtain infinite families of functionals, with each family corresponding to a specific choice of $\af$:
\ba
f_w^{\af}(z)=\sum_{n=1}^\infty f_n^{12,\af}(z) \, G_{2\af+2n}^{12}(w)\,.\label{eq:decompfa}
\ea
We find
\ba
f_{1}^{12,\af}(z)&=\frac{1}{z^{\af+2}}\\
f_{2}^{12,\af}(z)&=\frac{1}{z^{\af+2}}\left(\frac{1}{z}+\frac{\Delta_{12}^2-4(1+\af)^2}{12+8\af-2d}\right)\\
&\vdots\\
f_n^{12,\af}(z) &= \frac{q_n^{12,\af}(1/z)}{z^{\af+2}} \ , \\
&\vdots
\ea
with $q_n^{12,\af}$ (in spite of appearances) a polynomial of degree 
% $n$ 
$n-1$:
\unskip\footnote{
	A way to determine the general $n$ result is to note that
	\ba
	\frac{1}{z^{\frac{d+2}2}(1-z)^{\frac{2-d}2}}[C^{12}_w-C^{12}_z][z^{\frac{d+2}2}(1-z)^{\frac{2-d}2} f_w^{\af}(z)]=(\D_{12}^2-4\af^2) \left(\frac{w}{z}\right)^{1+\af}. \no
	\ea
	Plugging in the decomposition \rf{eq:decompfa} and \rf{F0} from the next section, this determines an equation for $f_{n}^{12,\af}$ which can be solved for.
}
\begin{multline}
q_n^{12,\af}(x)=
\frac{(-1)^{n+1} \G(2\af+n+1-\tfrac{d}{2}) \G(\tfrac{2\af+\D_{12} +2n}{2})\G(\tfrac{2\af -\D_{12}+2n}{2})}{\Gamma(n)\G(2\af+2n-\tfrac{d}{2})} \\
\qquad \qquad \qquad \times \,  {}_3\tilde{F}_2(1,1-n,2\af+n+1-\tfrac{d}{2};1+\af+\tfrac{\D_{12}}{2},1+\af-\tfrac{\D_{12}}{2}; x) \la{kern}
\end{multline}
Since they satisfy all the requirements set out in the previous subsection, these functionals lead to sum rules on the BOE. From the master functional action,
\ba
\Lambda_w^{12,\af}(\Delta)=G^{12}_{\Delta}(w)-\mathcal L_{\Delta}^{12,\af}(w)=\sum_{n=1}^\infty \theta_n^{12,\af}(\Delta) \, G_{2\af+2n}^{12}(w)\ , \la{eq:thma}
\ea
we find that the functionals corresponding to $f_n^{12,\af}$ are nothing but $\theta_n^{12,\af}$. Hence the sum rules associated to these functionals are precisely the ones postulated in Section \ref{Sdis},
\ba
\sum_{\Delta\geq 0} c_{\Delta} \, \theta_n^{12,\af}(\Delta)=0\,, \qquad n\geq 1\,.\la{sr}
\ea
 These functional actions can be determined from the relation of the $f_n^{12,\af}$ with the master functional given above in \rf{eq:thma}, or more directly by use of \rf{eq:basicfuncac}. From $\mathcal L_{2\af+2n}^{12,\af}(z)=0$ for $n\geq 1$ we find that the functionals $\theta_n^{12,\af}(\Delta)$ satisfy the duality properties:
\begin{equation}
\boxed{
\theta_{n}^{12,\af}(\Delta^{\af}_m)=\delta_{n,m}\,, \qquad \Delta^{\af}_m=2\af+2m\,, \qquad m\geq 1 \, . \la{dual}
}
\end{equation}
In a sense, these relations are the defining property of these functionals. They will be important for us later.

To conclude, note that the results of this subsection have established the implication ($\Rightarrow$) in equation \rf{eq:sumrules}. Let us now argue that ($\Leftarrow$) also holds, i.e.\ that validity of the sum rules implies locality of the form factor, provided that swapping holds. To see this, note that the kernels $z^{\af}f_n^{\af,12}(z)$ form a complete set of polynomials in $1/z$. Hence we have (using swapping)
\ba
&&\sum_{\Delta} c_{\Delta} \theta_n^{12,\af}(\Delta)&=0& \quad &\mbox{for all $n\geq 1$}
\\
&\Leftrightarrow &
\theta_n^{12,\af}[F]&=0& \quad &\mbox{for all $n\geq 1$}\\
&\Leftrightarrow& \int_0^1 \ud z\, z^{n} H(z)&=0& \quad &\mbox{for all $n\geq 0$} \label{eq:zn}
\ea
with
\ba
H(1/z)=\mathcal I_z\left[ z^{-\af} F(z)\right]\,.
\ea
The final line of \rf{eq:zn} implies that $H(z)$ must be the zero distribution, thus establishing locality of $F$.

A shortcoming of this argument is that it was necessary to take swapping as an assumption. A sufficient condition for this was polynomial boundedness \rf{bas} of $\tilde F(z)$ at $z=1$. It would be nicer if this could be established directly from absolute convergence of the sum rules, or perhaps by imposing extra regularity constraints on the BOE data. 

\subsubsection{Family relations}
\la{sec:family}
Before we conclude, let us discuss the relation between the families of functionals labeled by different $\af$. The basic observation is that
\ba
f_n^{12,\af}(z)=\frac{1}{z^{2+(\af-1)}} \left[\frac{1}{z} q_n^{12,\af}(1/z)\right]&=%z^{-2-\af} \left[\sum_{k=0}^{n+1} e_{n,k}^{12,\af+1} q_{k}^{12,\af}(1/z)\right]\\
\sum_{k=1}^{n+1} e_{n,k} f_k^{12,\af-1}(z)
\ea
for some coefficients $e_{n,k}$, since the $q_n$ are polynomials. To determine $e_{n,k}$, we act with both sides of this equality on boundary blocks with dimension $\Delta=2\af+2m$ and use the duality properties \rf{dual}. This sets all coefficients to zero except for two,
\ba
f_n^{12,\af}(z)&=f_{n+1}^{12,\af-1}(z)+\theta_n^{12,\af}(2\af) f_{1}^{12,\af-1}(z)\,,\\
\Rightarrow \qquad \theta_n^{12,\af}(\Delta)&=\theta_{n+1}^{12,\af-1}(\Delta)+\theta_n^{12,\af}(2\af)\theta_{1}^{12,\af-1}(\Delta) \label{eq:funcrels}
\ea
Using the explicit expressions for the functional kernels and demanding the right fall off at large $z$ constrains:
\ba
-\theta_n^{12, \tilde \a}(2\af)\equiv \frac{(-1)^{n} \left(\af-\frac{\Delta _{12}}{2}\right)_n \left(\af+\frac{\Delta _{12}}{2}\right)_n}{n! \left(2 \af-\frac{d}{2}+n\right)_n}\,. \la{enc}
\ea
This can be checked using the explicit expressions \rf{thr} for the functional actions given in the next section.
The relation \rf{eq:funcrels} between functional actions induces a similar relation between local blocks:
\ba
\mathcal L_{\Delta}^{12,\af}(z)&=\mathcal L_{\Delta}^{12,\af-1}(z)-\theta_1^{12,\af-1}(\Delta)\sum_{n=0}^\infty \theta_n^{12, \tilde \a}(2\af) \, G^{12}_{2\af+2n}(z)\\
&=\mathcal L_{\Delta}^{12,\af-1}(z)+\theta_1^{12,\af-1}(\Delta) \, z^{\af} \la{LBL}
\ea
The result for the infinite sum over blocks in the first line can be checked explicitly, and will be further justified in the next section.
The relation \rf{LBL} naturally suggests the following asymptotic of local blocks
\ba
\mathcal L_{\Delta}^{12,\af}(z)\underset{z\to \infty}\sim \theta_1^{12,\af-1}(\Delta) \, z^{\af}\,. \label{eq:asympL}
\ea
One can see that this is correct as follows. Since local blocks are themselves local, they can be bootstrapped by acting with functionals. Specifically, for consistency of the expansion \rf{lbe}, they should be bootstrappable with the $\theta^{12,\tilde\a}_n$ functionals. The only sufficiently suppressed asymptotic consistent with \rf{LBL} is \rf{eq:asympL}. One can thus understand \rf{LBL} as explaining how local blocks $\ml^{12,\tilde \a}_\D$ with more suppressed asymptotic can be obtained from $\ml^{12,\tilde \a-1}_\D$ by subtracting off the leading power. 

A final set of relations can be obtained as follows. Consider acting with the Casimir operator on the local block in the following way:
\ba
\frac{C^{12}_z-\lambda_{2\af}}{\lambda_{\Delta}-\lambda_{2\af}} \mathcal L_{\Delta}^{12,\af-1}(z)=G^{12}_{\Delta}(z)-\sum_{n=1}^\infty \left(\frac{\lambda_{2\af+2n}-\lambda_{2\af}}{\lambda_{\Delta}-\lambda_{2\af}}\right) \theta_{n+1}^{12,\af-1}(\Delta) G^{12}_{2\af+2n}(z) \ .
\ea
Notice that the right-hand side has the same kind of BOE as the local block with parameter $\af$. Furthermore, from the left-hand side and the asymptotics \reef{eq:asympL}, we also know that it scales as $z^{\af}$ for large $z$. This uniquely identifies the right-hand side as the local block with parameter $\af$, and so we discover that:
\ba
\theta_{n}^{12,\af}(\Delta)=\left(\frac{\lambda_{2\af+2n}-\lambda_{2\af}}{\lambda_{\Delta}-\lambda_{2\af}}\right) \theta_{n+1}^{12,\af-1}(\Delta) \ .
\ea
We can  combine this last result with \rf{eq:funcrels} to get
\ba
\theta_n^{12,\af}(\Delta)%&=e_{n-1}^{12,\af+1 } \left(\frac{\lambda_{\Delta}-\lambda_{2\af+2}}{\lambda_{\Delta}-\lambda_{2\af+2n }}\right) \theta_1^{12,\af}(\Delta)\\
&=\frac{(-1)^{n} \left(\af-\frac{\Delta _{12}}{2}\right)_n \left(\af+\frac{\Delta _{12}}{2}\right)_n}{n! \left(2 \af-\frac{d}{2}+n\right)_n}\left(\frac{\lambda_{2\af+2n}-\lambda_{2\af}}{\lambda_{\Delta}-\lambda_{2\af+2n }}\right) \theta_1^{12,\af-1}(\Delta) \ ,
\ea
which fixes the dependence of the functional actions on $n$. This is equivalent to the identity: 
\ba
\relax [C^{12}_z-\lambda_{\Delta}] \mathcal L^{12,\af}_{\Delta}(z)=\left(\Delta_{12}^2-4 \af^2\right)\theta_1^{12,\af-1}(\Delta) \, z^{\af+1} \ ,\la{lbc}
\ea
which is perfectly consistent with the asymptotics \rf{eq:asympL}.

\section{Applications}\la{WA}
In this section, we use the functionals constructed above to analytically bootstrap various examples of local form factors. We will see our results match up against various explicit computations for perturbative QFTs in AdS.

The functional kernels, actions and local blocks depend on the external dimensions $\D_1,\D_2$ only through their difference $\Delta_{12}$. Thus for simplicity, when $\Delta_1=\Delta_2$, we will often abbreviate $\theta_n^{12,\af}\to \theta_n^{\af}$ and so on.

\subsection{Bootstrapping GFF contact diagrams \label{Scont}}
We begin by considering the theory of a free 
scalar field $\Phi$ in AdS$_{d+1}$ with mass $m^2=\Df(\Df-d)$. Such a theory is dual to a Generalized Free Field (GFF) CFT on the AdS boundary with elementary field $\phi$. While the BOE of the field $\Phi$ itself is trivial in this case (it only contains $\phi$), this is not the case for composite operators. As a first application of our functionals we will show how to  bootstrap the form factor $\langle \Phi^2\, \phi \, \phi \rangle$. In fact it is easy to generalize our computation to $\langle \Phi^{2p} \, \phi^p \, \phi^p\rangle$ but we will keep $p=1$ for clarity of presentation.

\begin{figure} 
    \centering
    \includegraphics[width=6cm]{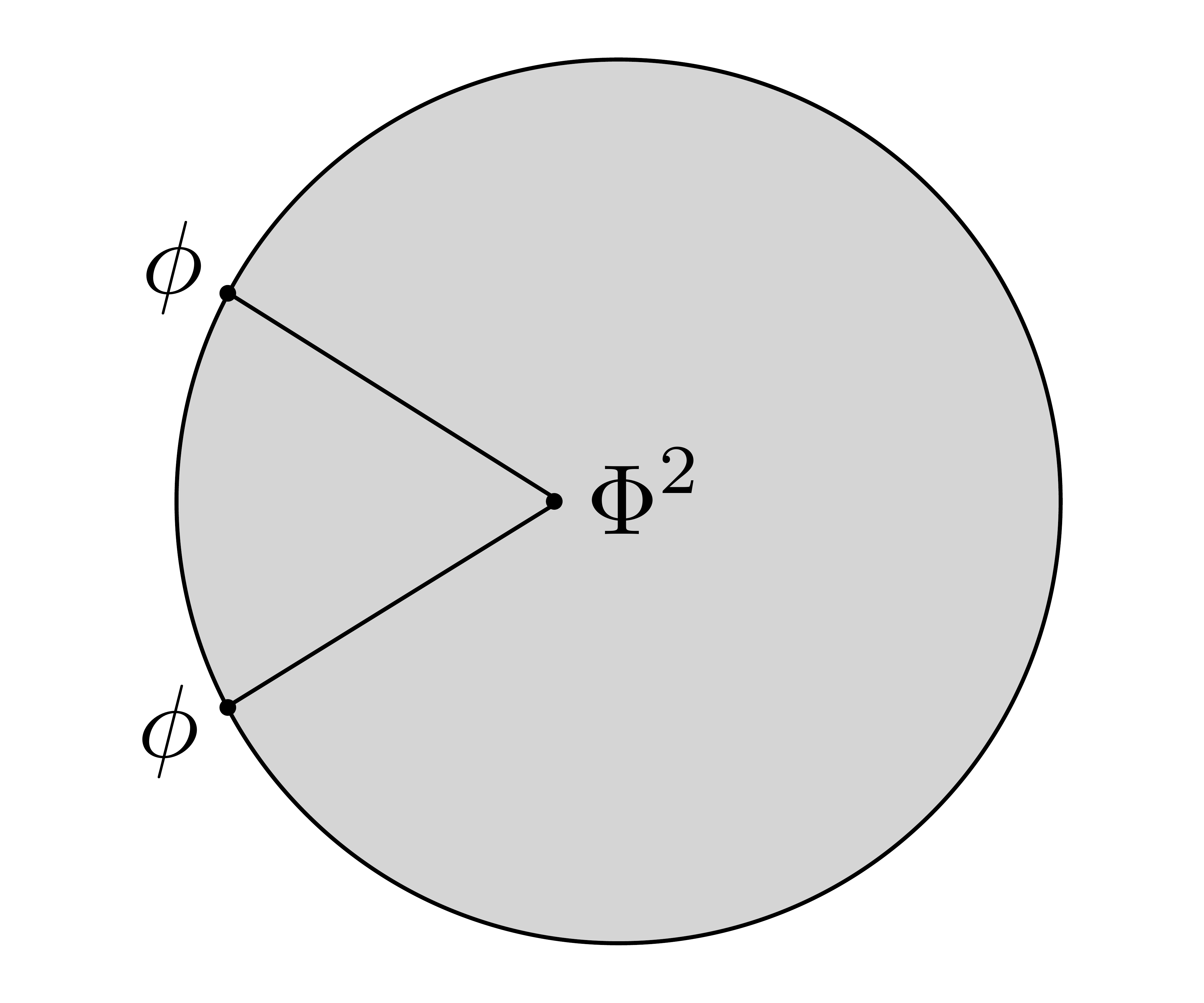}
    \hspace{1cm}
        \includegraphics[width=6cm]{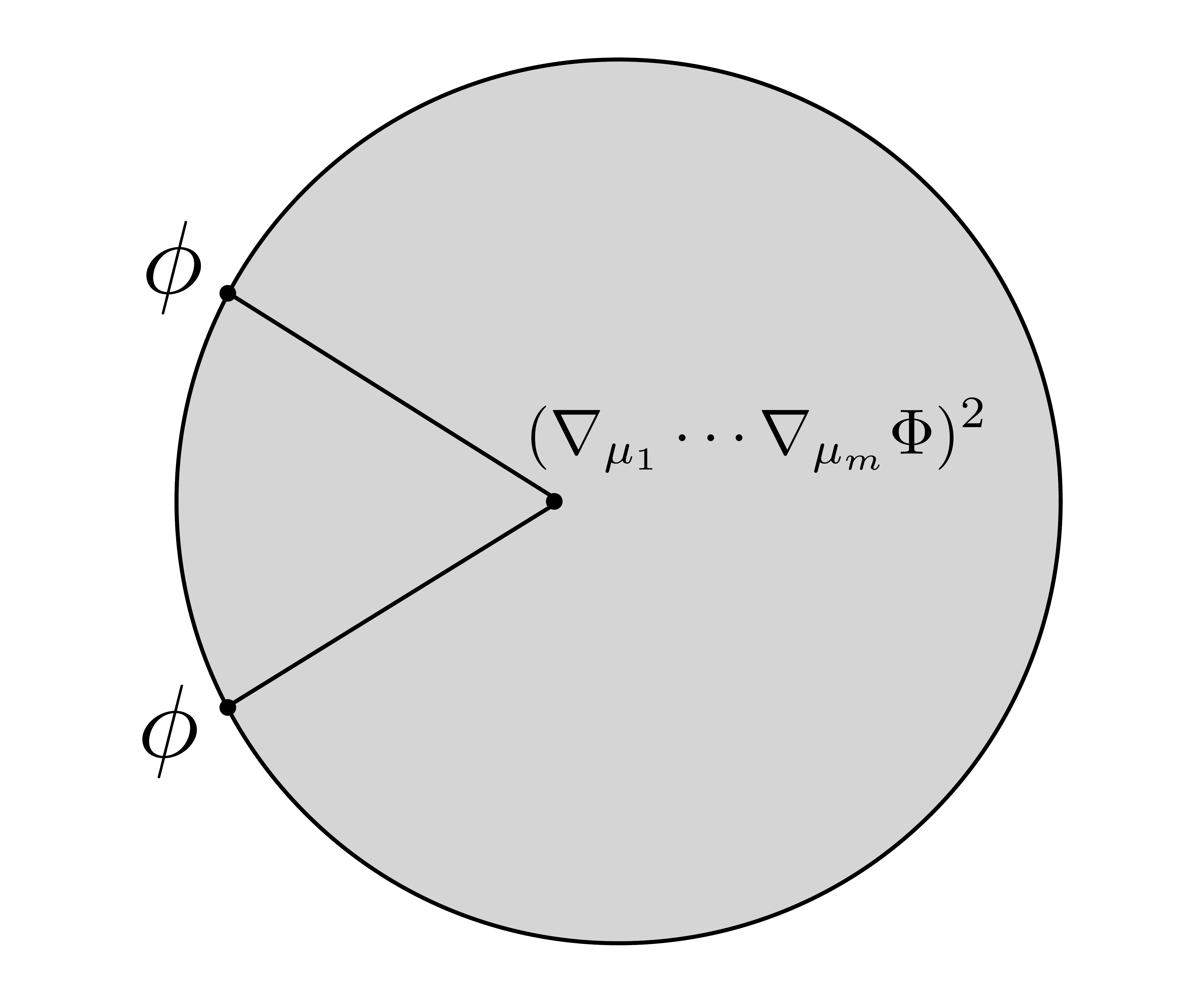}
    \caption{`Contact' Witten diagram in AdS. The lines denote free field propagators and the dots denote field insertions.
    }
    \label{fig:contact}
\end{figure}

Since we are dealing with a free theory, the operator $\Phi^2$ only couples to boundary operators which we denote as $(\phi^2)_n$. These operators have the schematic form $\phi\Box^n \phi$ and their scaling dimensions are $\D_n = 2\Df+2n$. It follows that the form factor must have a BOE of the form:
\ba
F(z)=\sum_{n=0}^\infty c_n \, G_{\Delta_n}(z)\,. \label{eq:BOEF0}
\ea
Indeed, a direct computation in AdS yields
\be
\langle \Phi^2(X) \phi(P_1)\phi(P_2)\rangle=\frac{1}{(-2P_1\cdot X)^\Df (-2P_2\cdot X)^\Df}\qquad \Rightarrow \quad F(z) =  z^{\Df} \,.\la{F0s}
\ee
As we will see below, this is consistent with the BOE given above. Given that the Witten diagram leading to this and other related computations involves no bulk exchanges, we call them `contact terms' (see figure \ref{fig:contact}).

Let us now bootstrap the same result using the locality constraints. We will assume:\footnote{Strictly speaking, what we should assume is the asymptotic of $\tilde F(z)$ at $z=1$ (defined in \rf{FTD}). In practice, in all the examples we have checked, this matches the large $z$ asymptotic of $F(z)$.}
\begin{itemize}
\item The {\em asymptotics} of $F(z):~\alpha_F=\Df$.
\item The \textit{spectrum}: the BOE only contains the operators with dimensions $\D_n = 2\Df+2n$.
\end{itemize}
To begin with, note that the assumption $\a_F = \Df$ implies we may only apply functionals with $\af>\Df-1$ to satisfy~\rf{constr}. Given the form of the BOE \rf{eq:BOEF0}, it is natural 
to set $\af=\Df$, so that the functionals are dual in the sense of \rf{dual} to the correct dimensions
in the BOE of $\Phi^2$.
As a result, the sum rules \rf{sr} are `diagonalized',
\be
0 = \sum_{\Delta\geq 0} c_{\Delta} \theta_n^{\Df}(\Delta)=\sum_{m=0}^\infty c_m \theta_n^{\Df}(2\Df+2m)=c_n + c_0 \, \theta_n^{\Df} (2\Df)  \ , \qquad n\geq 1 \ . \la{diag}
\ee
The coefficients $c_n = \mu_n \l_{n12}$ are thus uniquely fixed by locality. Using \rf{enc} we get:
\be
\frac{c_n}{c_0} = - \theta_n^{\Df} (2\Df) =  \frac{(-1)^{n}}{n!} \, \frac{(\Df)_n^2}{(2\Df-\frac d2+ n)_n} \ .\la{coef}
\ee
The overall factor $c_0$ is undetermined, as it corresponds to the freedom to rescale the bulk field $\Phi^2 \to \l \Phi^2$. Substituting these coefficients into the BOE and choosing $c_0=1$ reproduces~\rf{F0s}:
\be
F(z) = \sum_{n=0}^\infty \, \frac{(-1)^{n}}{n!} \, \frac{(\Df)_n^2}{(2\Df-\frac d2+ n)_n}\, G_{2\Df+2n}(z) = z^{\Df} \ . \la{F0}
\ee

An unsatisfactory feature of this computation was that the form factor's asymptotics $\aF=\Df$ were input by hand. To justify this assumption, we note that it is the simplest choice leading to a non-trivial solution. Indeed, for any $\aF<\Df$, we could have used functionals with $\af=\Df-1$. It is easy to check that, in that case, we would have found that all coefficients $c_n$ would have to be zero. 

Conversely, let us now examine what happens when we relax instead of tightening the asymptotics, keeping the assumed BOE spectrum $2\Df+2n$ the same.
We still want functionals dual to the correct spectrum, so we will choose:
\ba
\af=\Df+m \ , \qquad m\in \mathbb{N}.
\ea
Such functionals are suitable for bootstrapping form factors with $\aF<1+\af$. We now have the  duality conditions
\ba
\theta_n^{\Df+m}(2\Df+2m+2p)=\delta_{n,p}\ ,\qquad n,p\geq 1 \ .
\ea
Effectively, we see that, every time we increase $m$, we lose constraints. This is manifest in the identities~\rf{eq:funcrels}, which show we can increase $m$ by one unit by sacrificing a functional. Since we have lost constraints, the set of form factors compatible with our assumptions is now larger. The sum rules now give:
\ba
c_{m+n}+\sum_{k=0}^{m} c_k \, \theta_n^{\Df+m}(2\Df+2k)=0 \ , \qquad n\geq 1\ . \la{srn}
\ea
and hence, solutions are labelled by $m$ additional parameters $c_1\,, \ldots, c_m$. Of course, the solutions with a given value of $m$ include all of those with lower values of $m$. 

To understand the physical origin of these extra solutions, let us note that our BOE is compatible not only with the form factor of $\Phi^2$, but more generally with 
\ba
\langle (\nabla_{\mu_1}\ldots \nabla_{\mu_m} \Phi)^2 \, \phi \,  \phi\rangle \label{eq:c2m}
\ea
To show that these indeed correspond to the solutions above, consider the following change of basis\footnote{After using the  bulk equation of motion  $(\nabla^2_{\mbox{\tiny AdS}}-\l_{\D_\p})\Phi=0$ to remove terms of the form $\nabla^2_{\mbox{\tiny AdS}} \Phi$,  we see that \rf{opr} are  indeed  linear combinations of the  form factors \rf{eq:c2m},
$$
F^{(m)} = \frac{(-2)^{m}}{\prod_{l=0}^{m-1} (2\D_\p+2l)^2} \langle  (\nabla_{\m_1}\cdots \nabla_{\m_m}\Phi)^2 \,  \phi \, \phi \rangle + \sum_{l=1}^{m-1} c_{ml} F^{(l)} \ ,
$$
for some coefficients $c_{ml}$.}
\ba
F^{(m)}(X,P_1,P_2)&:= \Big\langle \left[\left(\prod_{l=1}^{m} \frac{\l_{\D_l}-\nabla^2_{\mbox{\tiny AdS}}}{(2\D_\p+2l)^2} \right) 
\Phi^2(X)  \right] \, \phi(P_1)\, \phi(P_2) \Big\rangle\\
\Rightarrow\qquad F^{(m)}(z)& = \left(\prod_{l=1}^{m} \frac{\l_{\D_l}-C^{12}_z }{(2\D_\p+2 l)^2} \right) F^{(0)}(z) \ , \qquad F^{(0)}=z^{\Df}\ . \la{opr}
\ea
Since $C^{12}_z$ is the Casimir operator (of which boundary blocks are eigenfunctions, cf.\ \rf{eq:cas}), 
we have
\ba
F^{(m)}(z)=\sum_{k=m}^\infty c_k^{(m)} \, G_{2\Df+2k}(z)
\ea
for some coefficients $c_k^{(m)}$. Since the BOE of $F^{(m)}(z)$ has $c_0=c_1=\ldots c_{m-1}=0$, the sum rules \rf{srn} predict
\ba
c_{n}^{(m)}\bigg|_{\Df}=c_{n}^{(0)}\bigg|_{\Df\to \Df+m}
\ea
and hence imply $F^{(m)}(z)=z^{\Df+m}$. This is indeed correct, as follows from direct computation of \reef{opr}.

An alternative way to derive these results is to use the local block expansion \eqref{eq:localbs2}. Let us make the same assumptions on the BOE spectrum and set $F(z)=O(z^{\Df+m})$ with arbitrary $m$. Then:
\ba
F(z)=\sum_{n=0}^\infty c_n G_{2\Df+2n}(z)=\sum_{n=0}^\infty c_n \mathcal L^{\Df+m}_{2\Df+2n}(z)=\sum_{n=0}^m c_n \mathcal L_{2\Df+2n}^{\Df+m}(z)
\ea
where we used $\mathcal L^{\af+m}_{2\af+2n}(z)=0$ for $n>m$. Hence the form factors become identified with simple combinations of local blocks. 
In particular, the form factors $F^{(m)}=z^{\Df+m}$ defined above correspond to single local blocks,
\be
F^{(m)}(z)=z^{\Df+m} =  \ml_{2\Df+2m}^{\Df+m} = G_{2\Df+2m}(z)-\sum_{n=1}^\infty \theta^{\Df+m}_n(2\Df+2m) G_{2\Df+2m+2n}(z)\ , 
\ee
where we have used \rf{lbe}. This boundary block expansion indeed matches the computation from sum rules above.
In fact, we can actually directly compute these local blocks as functions, rather than as boundary block expansions. Using \eqref{eq:replocal1} and \eqref{eq:idblock},   we find:
\begin{align}
\mathcal L_{\Delta}^\Df(z)=\frac{\sin\left[\frac{\pi}2(\Delta-2\Df)\right]}{\pi} \int_{-\infty}^0 \ud w &\frac{(-z/w)^{\Df+1}}{z-w} \tilde G_{\Delta}(\mbox{$\frac{w}{w-1}$}) \la{intf}\\
&\underset{\Delta\to 2\Df}\sim z^{\Df} \frac{2 \sin\left[\frac{\pi}2(\Delta-2\Df)\right]}{\pi(\Delta-2\Df)}\underset{\Delta= 2\Df}\to z^{\Df}
\end{align}
What made an exact computation possible was the fact that, for this particular value of $\Delta$, the integral is dominated by the region $w\sim 0$ where the integrand simplifies. It is also easy, by subtracting powers from the integrand and adding back their integral, to analytically extend the integral formula \rf{intf} for the local block. From the resulting formula, one can find exact expressions for $\Delta=2\af-2p$ for any integer $p$:
\ba
\mathcal L^{\af}_{2\af-2p}(z)=\sum_{k=0}^{p} b_k \, z^{\af-k} \ ,
\ea
where the coefficients $b_k$ can be determined exactly. Here we just point out that this is perfectly consistent with our contact term computations.

\subsection{Local blocks from AdS Witten diagrams \label{localS}}
The local blocks $\ml_\D^{12,\tilde \a}$ developed in Section \ref{Sdis} are, in particular, local form factors. 
Thus one may wonder if there a local QFT in AdS that realises them as its form factors. Given that their expansion \rf{lbe} in boundary blocks  contains  $G_{\D}^{12}$ and $G_{2\tilde \a+2n}^{12}$ ($n\geq 1$), it is natural to guess that this is the theory of three scalars $\Phi_1$, $\Phi_2$ and $\Psi_\D$ in AdS (corresponding to boundary fields of dimension $\D_1$, $\D_2$ and $\D$) coupled through an interaction vertex $\Psi_\D \Phi_1 \Phi_2 $. The goal of this section is to show that this guess is correct. 
\begin{figure}
    \centering
    \includegraphics[width=7cm]{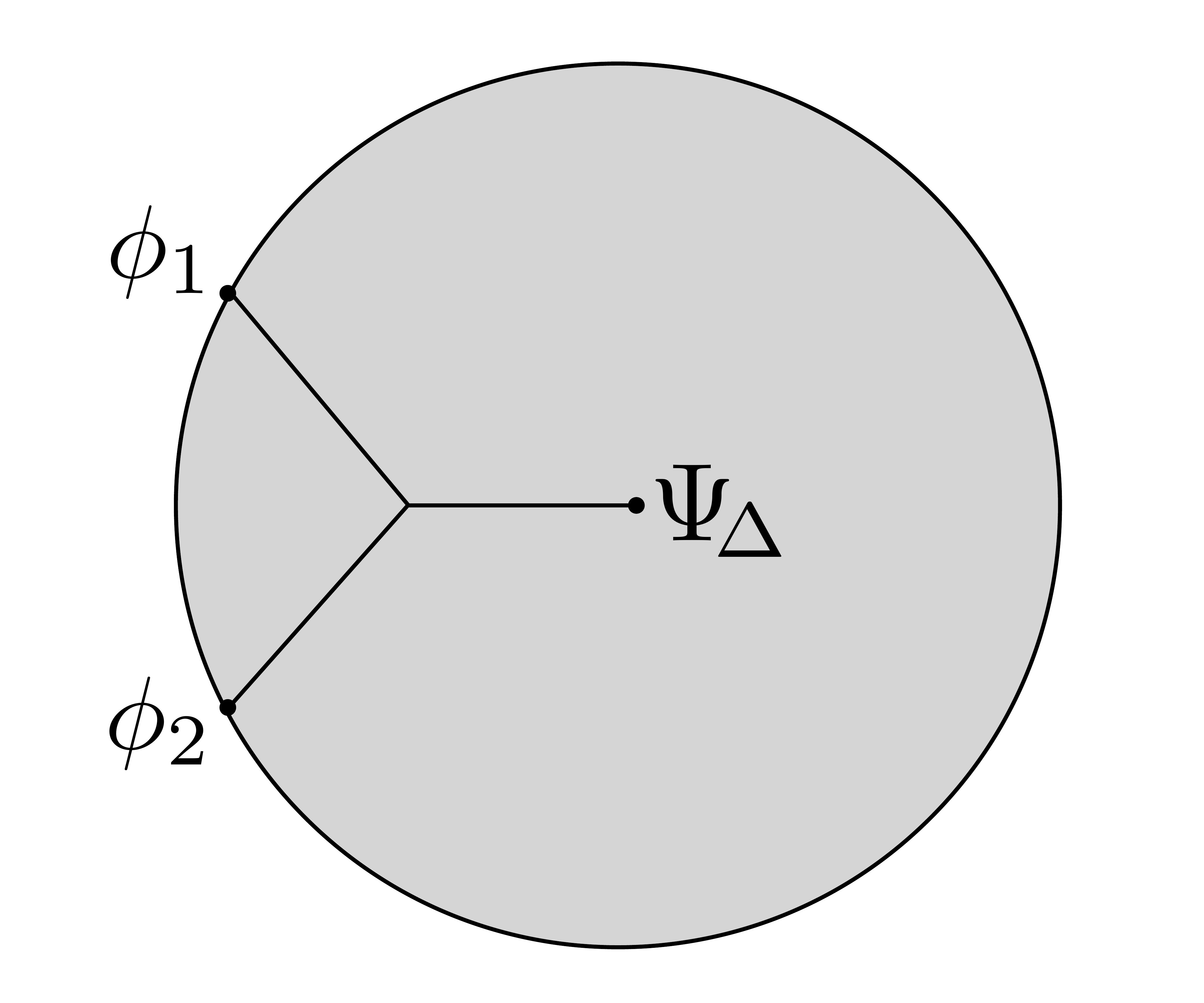}
    \caption{`Exchange' Witten diagram in AdS.
    }
    \label{fig:exch}
\end{figure}

Consider the form factor 
$E_\D^{12} = \langle \Psi_\D \, \phi_1 \, \phi_2 \rangle$ in such a theory,
which is given at tree level by a single `exchange' Witten diagram
(see figure \ref{fig:exch}),
\begin{align}
&E_\D^{12}(z)  
= (-2 P_{12})^{(\D_1+\D_2)/2}\left(\frac{P_1 \cdot X}{P_2 \cdot X}\right)^{\frac{\D_{12}}2}\int_{\ttext{AdS}}   \ud {X'} \, \frac{P^{BB}_{\Delta}(X,X')}{(-2P_1\cdot X')^{\D_1}(-2P_2\cdot X')^{\D_2}} \ . \la{ES}
\end{align}
Here we are using the embedding space formalism introduced in Section \ref{AdS}. We use the following expression for the bulk-to-bulk propagator \cite{Penedones:2010ue}:
\begin{align}
&P^{BB}_{\Delta}(X,{X'})=\int_{-i\infty}^{i\infty} \frac{\ud c}{2\pi i} f(c) \int_{\mathbb R^{d}}\, \ud Q \frac{1}{(-2Q\cdot X')^{\frac d2+c}(-2Q\cdot X)^{\frac d2-c}}\ ,\\
&f(c)=\frac{1}{2\pi^d} \frac{\Gamma(\frac d2+c)\Gamma(\frac d2-c)}{\Gamma(c)\Gamma(-c)} \frac{1}{\left(\Delta-\frac d2\right)^2-c^2}\ .
\end{align}
The  integrals over $X'$ and $Q$ may be performed using the integral identities stated in Appendix~\ref{inti}.
As a result we find 
\begin{align}
&E_\D^{12}(z)=  \int_{-i\infty}^{i\infty} \frac{\ud c}{2\pi i} \, \h_\D^{12}(c) \, G_{\frac d2-c}^{12}(z) \ ,  \la{lb} \\ 
&\h_\D^{12}(c)= \frac{\Gamma\left(\frac{\D_1+\D_2-d/2+c}2\right)\Gamma\left(\frac{\D_1+\D_2-d/2-c}2\right) \Gamma\left(\frac{d/2-c+\D_{12}}2\right)\Gamma\left(\frac{d/2-c-\D_{12}}2\right)}{2 \, \Gamma(-c)\Gamma(\D_1)\Gamma(\D_2)\left[(\Delta-\frac d2)^2-c^2\right]} \no \ .
\end{align}
Closing the contour on the left-hand side, we pick up the pole at $c=\frac d2-\Delta$ and the series of poles at $c=\frac d2-(\D_1+\D_2+2n)$ to obtain
\begin{align}
    E_\D^{12}(z)=a^{12}(\D)\, G_{\D}^{12}(z) + \sum_{n=0}^\infty a_{n}^{12}(\D)\, G_{\D_1+\D_2+2n}^{12}(z) \ , \la{Ee}
\end{align}
where 
\be
a^{12}(\D) = \underset{c=\tfrac{d}{2}-\D}{\text{Res}} \h_\D^{12}(c)\ , \qquad a^{12}_n(\D) = \underset{c=\tfrac{d}{2}-(\D_1+\D_2+2n)}{\text{Res}} \h_\D^{12}(c) \la{ade} \ .
\ee
Now comparing \rf{Ee} with the expression \rf{lbe}, we assert that $E_\D^{12}$ is proportional to a local block:
\ba
E_\D^{12} = a^{12}(\D) \, \ml_\D^{12, \tilde \a}   \qquad  \mbox{with}\quad  \tilde\a = \ha(\D_1+\D_2)-1 \ .\label{EtoL}
\ea
This identification of the exchange diagram as a local block is correct because: 
(i) it is local, since it is the tree-level contribution to a local form factor, 
(ii) it contains only the blocks $G_\D^{12}$ and $G_{2\tilde \a+2n}^{12}$ ($n\geq 1$),  and (iii) it has  large $z$ asymptotic $\sim$$z^{\tilde \a}$.\footnote{The asymptotic behaviour $E_\D^{12} \sim z^{(\D_1+\D_2)/2-1}$ follows from the standard fact that $(C^{12}_z-\l_\D)E_\D^{12}$ equals a contact term $z^{(\D_1+\D_2)/2}$ (see eq.\ \rf{ice} below).}.  There is only one such function (up to rescaling), as can be seen by applying the functionals $\theta^{12,\tilde \a}$. 

Thus the coefficients in \rf{Ee} are related to the functional actions $\theta_n^{12,\tilde \a} ( \D)$ defined in Section \ref{SR} (which are the coefficients for the local block expanded in boundary blocks),
\begin{align}
\theta_n^{12,\tilde \a} ( \D)= - \frac{a_{n-1}^{12}(\D)}{a^{12}(\D)}   \qquad \quad  (\tilde\a = \ha(\D_1+\D_2)-1)  \ .\la{thq}
\end{align}
Explicitly evaluating these residues of $h_\D^{12}$, we find the expressions \rf{thr} which we repeat here
\begin{align}
\theta_n^{12,\tilde \a} ( \D) 
&=\frac{4(-1)^n}{(n-1)!}\frac{ 1}{(\D-2\tilde \a-2n)(\D+2\tilde \a+2n-d)} \la{thr2}\\
&\qquad \qquad   \times \frac{\G(\tilde \a+n+ \tfrac{\D_{12}}{2})\G(\tilde \a+n- \tfrac{\D_{12}}{2})\G(\D+1-\tfrac{d}{2})\G(2 \tilde \a +n+1-\tfrac{d}{2})}{\G(\tfrac{\D}{2}+\tfrac{\D_{12}}{2})\G(\tfrac{\D}{2}-\tfrac{\D_{12}}{2})\G(\tilde \a-\tfrac{\D}{2}+1)\G(\tilde \a +\tfrac{\D}{2}+1-\tfrac{d}{2})\G(2 \tilde \a +2n-\tfrac{d}{2})} \ . \no
\end{align}
As a cross-check, this expression satisfies the identities derived in subsection \ref{sec:family}. In particular, the identity \rf{lbc} describing the action of the  Casimir operator on a local block translates to a relation between AdS exchange and contact diagrams: 
\unskip\footnote{Note that consistency requires a particular relation between $\theta_1^{12,\af-1}$ and $a^{12}(\Delta)$:
\be
a^{12}(\D) = \left( \frac{1}{4\tilde\a^2-\D_{12}^2} \right) \frac{1}{\theta_1^{12,\tilde \a-1}(\D)}  \qquad \quad  (\tilde\a = \ha(\D_1+\D_2)-1) \ . \no
\ee
which indeed holds, consistently with equations \rf{thr}, \rf{EtoL} and \rf{lbc}}
\be
(C^{12}_z- \l_\D) E_\D^{12}(z) = - \, C^{12}(z) \ , \qquad\qquad  C^{12}(z) = z^{(\D_1+\D_2)/2} \ . \la{ice}
\ee
As we explained previously, the action of the Casimir is the same as that of the AdS Laplacian on the bulk field. The combination $(C_2- \l_\D)$ is then $(\nabla^2_{\mbox{\tiny AdS}}-m_{\Delta}^2)$, which reduces the bulk-to-bulk propagator to a delta-function, thus turning the exchange diagram into a contact diagram \cite{DHoker:1999mqo}.

\subsection{$\Phi^4$ interactions \la{P4}}
\begin{figure}
    \centering
    \includegraphics[width=7cm]{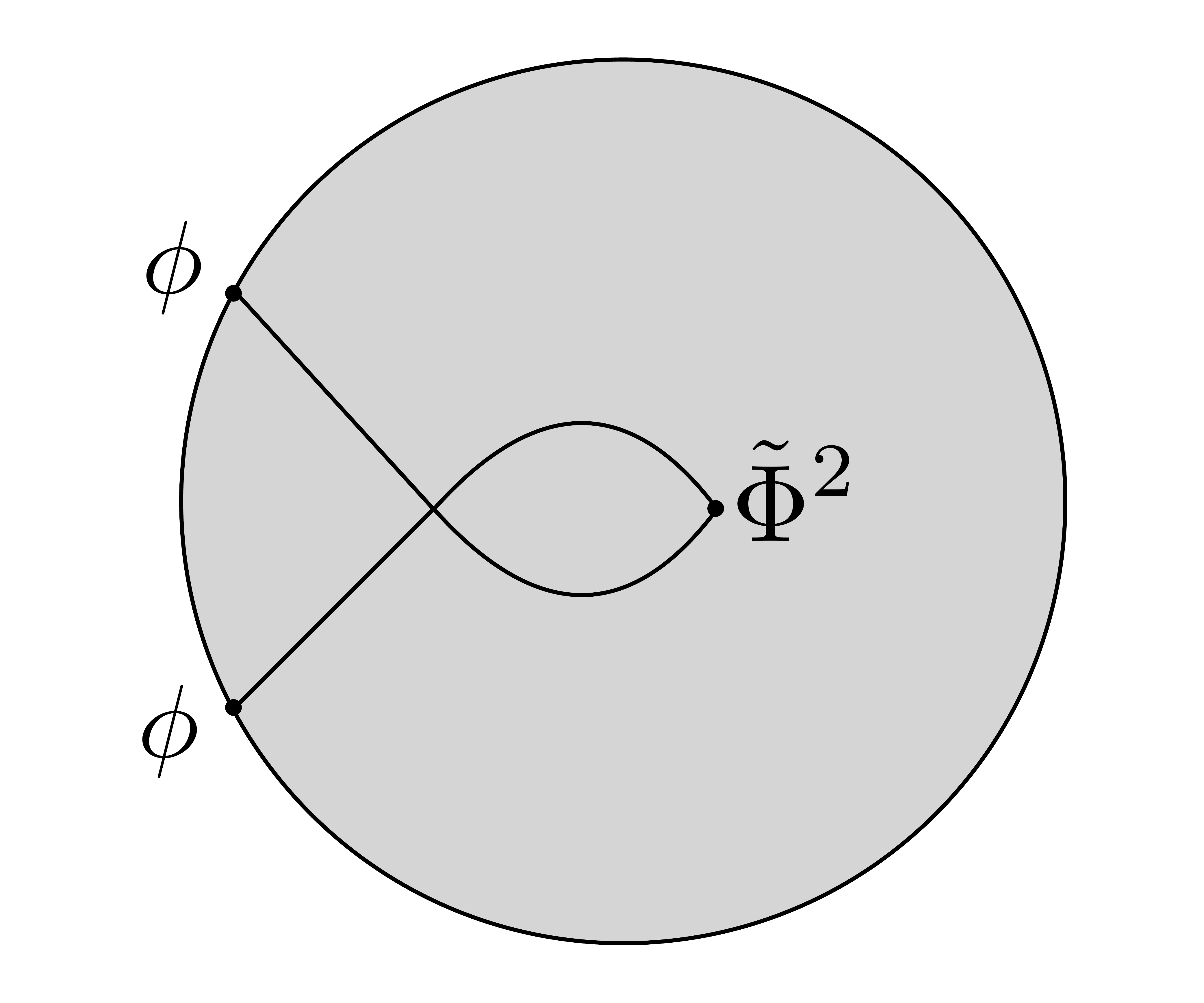}
    \caption{The Witten diagram contributing to the form factor $\langle \tilde \Phi^2 \, \phi\, \phi \rangle$ at leading order $\mathcal O(g)$.
    }
    \label{fig:phi4}
\end{figure}

As our next example, we consider a theory of two initially free scalar fields $\Phi$ and $\tilde \Phi$ in AdS, corresponding to dual bulk operators $\phi, \tilde \phi$ with dimensions $\Delta$ and $\tilde \Delta$ respectively. The fields are then coupled together via a quartic interaction of the form 
\ba
\mathcal L_{\mbox{\tiny int}} \sim 
g\, \Phi^2 \tilde 
\Phi^2.
\ea
Here we will show how to bootstrap the leading order result for the form factor $F=\langle  \tilde \Phi^2 \, \phi \, \phi \rangle$. The relevant $O(g)$ diagram is shown in figure \ref{fig:phi4}. To leading order in $g$, the BOE of $\tilde \Phi^2$ contains two towers of states,
\be
F(z) = \sum_{m=0}^\infty c_m \, G_{2\D+2m}(z) + \sum_{n=0}^\infty  d_n \, G_{2\tilde \D+2n}(z)  \ ,\la{bb4}
\ee
corresponding to `double trace' operators $(\phi^2)_n$ and $(\tilde \phi^2)_n$.
In the above we should have in mind that
\ba
c_n=\underbrace{\mu^{\tilde \Phi^2}_{(\phi^2)_n}}_{O(g)}  \underbrace{\lambda^{\phi^2}_{(\phi^2)_n}}_{O(1)}\,, \qquad d_n=\underbrace{\mu^{\tilde \Phi^2}_{(\tilde \phi^2)_n}}_{O(1)}  \underbrace{\lambda^{\phi^2}_{(\tilde \phi^2)_n}}_{O(g)} \ \ .
\ea
At this order in $g$, scaling dimensions are unmodified and no other operators appear. 
Our goal will be to determine $c_n$ so as to read off the new $\mu$ couplings at $O(g)$, as functions of the explicitly known coefficients $d_n$ (see Appendix \ref{supp}). Let us apply to $F$ a complete set of functionals $\theta_{n}^{\tilde \a}$; the natural choice is take $\tilde \a = \D -1$ so that the sum rules determine:
\be \begin{split}
&c_m=  -  \sum_{n=0}^\infty d_n \, \theta_{m+1}^{\D-1}(2\tilde \D+2n) \ , \qquad m\geq 0   \ .\la{scd}
\end{split} \ee
However, there is one caveat. In principle we need to establish that the asymptotics of the form factor to ensure the applicability of these functionals. To estimate the behaviour of $F(z)$ at large $z$, we recall that this is a high-energy limit where AdS space becomes effectively flat and particles  massless. From the momentum-space form factor we get
\ba
\mathcal F(p)\sim \int d^{d+1} p' \frac{1}{(p'+p)^2  p'^2} \underset{p\to \infty}\sim |p|^{d-3} \ .
\ea
Given that for contact terms we have $\mathcal F(p)\sim p^{2k}$, while in position space $ F(z)\sim z^{\Delta+k}$, we guess here
\ba
 F(z)\underset{z\to \infty}= O(z^{\frac{d-3}2+\Delta})\,.
\ea
\sloppy Thus we expect to require $d<3$ for validity of the sum rules. In fact we find by plugging in the explicit values of $d_n$ in Appendix \ref{supp}
that the summand in \rf{scd} goes like \mbox{$d_n \theta_{m+1}^{\D-1}(2\tilde \D+2n) \sim n^{d-4}$} for large $n$, so that the sums indeed converge only for $d<3$. 

The reason for this bound is related to the fact that the quartic interaction we have chosen is a relevant term (from a $d+1$ dimensional perspective) only for $d<3$. In the form factor computation, this manifests as the fact that, for $d<3$, the large $z$ asymptotics of the form factor are strictly weaker than those of a contact term. 
For higher $d$ this is no longer the case, so we must allow for a wider set of solutions, which may, in particular, have all $d_n$ switched off. Equivalently, the bulk field can always be shifted by $\tilde \Phi^2 \to \tilde \Phi^2 +c\, g\, \Phi^2$, shifting the form factor by a contact term; this contact term becomes indistinguishable from our desired form factor at the level of asymptotics as $d$ passes above $3$. In practice, we can proceed by assuming $d<3$ but our final results hold for any $d$ by analytic continuation. The net result is that the sum rules \rf{scd} determine the coefficients $c_n$ from the known ones $d_n$, bypassing any need for perturbative computations. Their explicit form is given in Appendix \ref{supp}.

Note that an alternative way to arrive at these results is to make use of of local blocks. Locality is equivalent to the expansion \rf{eq:localbs} in local blocks, 
\be
F = \sum_{m=0}^\infty c_m \, G_{2\D+2m}(z) + \sum_{n=0}^\infty  d_n \, G_{2\tilde \D+2n}(z) =  
\sum_{n=0}^\infty  d_n \, \ml_{2\tilde \D+2n}^{\D-1}(z) \ , \la{lb4}
\ee
where we have used the fact \rf{vl} that $\ml_{2\D+2m}^{\D-1}(z)=0$ for $m\geq 0$. 
Re-expanding the local blocks in boundary blocks as in \rf{lbe}, the coefficients $c_{m}$ are evidently determined in terms of the others, and will indeed be given by the sum rules \rf{scd}.

\bigskip
Let us now check the prediction of our sum rules by explicitly computing the AdS form factor in perturbation theory.
The leading $O (g)$ contribution corresponding to the diagram in figure \ref{fig:phi4} is given by
\be
F(z) =g \, (-2 P_{12})^{\D}\int_{AdS} dX' \frac{P^{BB}_{\tilde \D}(X',X)^2}{(-2P_1\cdot X')^{\D} (-2P_2\cdot X')^{\D}} \ . \la{p01}
\ee
The product of two bulk-to-bulk propagators between the same two points can be expanded as a sum of single propagators \cite{Fitzpatrick:2011hu},
\be
P^{BB}_{\D_a}(X',X) P^{BB}_{\D_b}(X',X) = \sum_{n=0}^\infty a_n \, P^{BB}_{\D_a+\D_b+2n}(X',X) \la{decomp} \ ,
\ee
where the coefficients $a_n$ are given explicitly in Appendix \ref{supp}. Using this decomposition with $\D_a=\D_b=\tilde \D$ in \rf{p01}, we get
\be
F(z)= (-2 P_{12})^{\D} \int_{\rm AdS} dX' \sum_{n=0}^\infty a_n \frac{P^{BB}_{2\tilde \D+2n}(X',X)}{(-2 P_1\cdot X')^{\D} (-2 P_2\cdot X')^{\D}}  \ .
\ee
Following the same argument as in Section \ref{localS}, we obtain
\begin{align}
F(z)&= \sum_{n=0}^\infty a_n \int _{-i \infty}^{i \infty} \frac{dc}{2\pi i}  h_{2\tilde \Delta+2n}(c) G_{\tfrac{d}{2}-c}(z) \ , \la{F3} \\ 
&= \sum_{n=0}^\infty a_n \,   E_{2\tilde \Delta+2n}(z)  \ , \la{F4}
\end{align}
where we denote
\be
h_{2\tilde \Delta+2n}(c) \equiv h_{2\tilde \Delta+2n}^{\D\D}(c) \ , \qquad \qquad  E_{2\D_1+2n}\equiv E_{2\tilde \Delta+2n}^{\D\D} \ , 
\ee
and $h_{2\tilde \D+2n}^{\D\D}(c)$, $E_{2\tilde \D+2n}^{\D\D}$ are as defined in Section \ref{localS}.

The form factor \rf{F4} is just a sum of the exchange diagrams \rf{lb} studied above. Using the fact that these exchange diagrams are proportional to local blocks \rf{EtoL}, we can then write the form factor in the manifestly local form \rf{lb4}: as a sum of local blocks
\begin{align}
&F(z)= \sum_{n=0}^\infty d_n \, \ml_{2\tilde \D+2n}^{\D-1}(z) \ , \la{lex} \qquad\qquad   d_n =  a_n \, a^{\D\D}(2\tilde\D+2n)   \ ,
\end{align}
where $a^{\D\D}(2\tilde\D+2n)$ were defined in \rf{ade}.
One can check that the values of $d_n$ resulting from \rf{lex} match those stated in \rf{7sol}.

As discussed above, the local block expansion \rf{lex} can be converted to a boundary block expansion by expanding the local blocks with the formula \rf{lbe}. The same is achieved by simply commuting the order of summation and integration in \rf{F3},
\be
F=  \int _{-i \infty}^{i \infty} \frac{dc}{2\pi i} \left[\sum_{n=0}^\infty a_n h_{2\tilde \Delta+2n}(c)\right] G_{\frac{d}{2}-c}(z)
\ee
The function in square brackets has poles at $c=\tfrac{d}{2}-(2\D+2m)$ and $c=\tfrac{d}{2}-(2\tilde \D+2n)$ for each $m$ and $n$. Closing the contour on the left, we obtain a boundary block expansion of precisely the form \rf{bb4}. The coefficients $d_n$ are, of course, the same as those in the local block expansion \rf{lex}. The remaining tower of coefficients, $c_{m}$ precisely match the values \rf{scd} predicted by the sum rules (stated in App.\ \ref{supp}).

To conclude, let us briefly consider the case where $\tilde \D=\D$ to study the form factor $\langle \Phi^2 \, \phi \, \phi \rangle$ for a single scalar field with $\frac{g}{4!}\Phi^4$ interaction in AdS and mass $m^2 = \D(\D-d)$. Taking the limit $\tilde \Delta \to \Delta$ of our previous computation determines the leading correction $F^{(1)}$ to the form factor $F^{(0)}=z^{\D}$ of the free theory (see eq.\ \rf{F0}):
\ba
  F^{(0)}(z) + g \, F^{(1)}(z)\sim \sum_{n=0}^\infty \left(c_n^{(0)} + g \, c_n^{(1)} \right) \,  G_{2\D+2n+ g \, \g_n} \ . \la{cF}\\
 \ea
In this limit, the sum rules determine the corrections $c_{n}^{(1)}$ to the coefficients in \rf{cF} in terms of the anomalous dimensions $\g_n$.
One finds that the coefficients $c_m$,$d_n$ diverge in this limit $\tilde \D \to \D$, but these  divergences cancel between the two towers:
\begin{align}
&c_n = \frac{u_n}{\D-\tilde \D} + v_n + \ldots \ , \qquad d_n = - \frac{u_n}{\D- \tilde \D} + w_n  \ ,\\
&F^{(1)}(z) = \sum_{n=0}^\infty\big[ 2\, u_n \, \partial_\D G_{2\D+2n} + (v_n+w_n) \, G_{2\D+2n}\big] \ .
\end{align}
In particular we can read off the anomalous dimensions,
\begin{align}
&\g_n = \frac{2u_n}{c_n^{(0)}} = \frac{2^{-3-d}\, \pi^{-\tfrac{d}{2}}\, \G(\tfrac d2+n)\G(\D+n)\G(\ha-\tfrac d2+\D+n)\G(-\tfrac d2+2\D+n)}{n! \ \G(\tfrac{d}{2})\G(\ha+\D+n)\G(1-\tfrac{d}{2}+\D+n)\G(1-d+2\D+n)} \ , \la{anom4}
\end{align}
which match the known result \cite{Fitzpatrick:2010zm}. 

\section{Locality constraints on CFT data}
\label{loccft}
One way of viewing the set of locality constraints \rf{lboots} is as a rulebook for re-constructing local bulk fields from boundary data.  However, it can also be seen as a set of constraints on the CFT data in order for it to be consistent with the existence of a local AdS bulk. The central idea is that, fixing a bulk field and considering several different choices of boundary operators, one can effectively `eliminate' the BOE coefficients $\mu_\D$ from the locality constraints, yielding relations purely among the CFT structure constants $\l_\D^{12} := \l_{\D}^{\D_1\D_2}$. 
In this section, we begin to explore such relations in a very simple context, namely the GFF CFT, exploiting the functional machinery built up in previous sections.

\subsection{Eliminating the BOE coefficients}
For a given bulk field $\Psi$, there is a list of locality constraints indexed by the external operators ${\mathcal O}_{\D_1},{\mathcal O}_{\D_2}$,
\be
\mathcal I_z\left[ \, \sum_n\mu_n^\Psi \,  \l_{\D_n}^{12} \,  G_{\D_n}^{12} (z)\right] = 0 \ , \qquad \mbox{for}\ z\geq 1\ .
\ee
Crucially, the BOE coefficients $\mu_n^\Psi$ are the \textit{same} for each choice of external operators.
Let us assume the BOE spectrum to be $\D_n=2\b+2n$ for some parameter $\b$.
Applying the dual set of functionals $\theta_n^{12,\b}$,  the combinations $\mu^\Psi_n \, \l_{\D_n}^{12}$ may be exactly bootstrapped up to an overall factor $\mu_0^\Psi \, \l_{\D_0}^{12}$,
\be
\frac{\mu_n^\Psi \, \l_{\D_n}^{12}}{\mu_0^\Psi \, \l_{\D_0}^{12}} = -  \theta_n^{12,\b}(2\b)  \ . \la{fe}
\ee
More precisely, this is true only if we assume that the form factor has asymptotic behaviour 
\be
F^\Psi_{\cO_1\cO_2}(z) \underset{z\to\infty}{\sim} z^{\b+1-\epsilon}  \qquad  (\epsilon>0)\ , \la{aba}
\ee
so that the functionals $\theta_n^{12,\b}$ satisfy the constraint \rf{constr} (we will discuss more general cases below). It follows from \rf{fe} that the combinations $- \tfrac{\mu_n^\Psi}{\mu_0^\Psi} = \tfrac{ \l_{\D_0}^{12}}{ \l_{\D_n}^{12}} \theta_n^{12,\b}(2\b)$ are independent of the choice of operators $1,2$, as long as the asymptotic constraint above continues to be satisfied. Therefore, for any such pair, we have
\be
\frac{\l_{\D_n}^{12}}{\l_{\D_0}^{12}} = \frac{\theta_n^{12,\b}(2\b)}{\theta_n^{\widehat{12},\b}(2\b)}  \ \frac{\l_{\D_n}^{\widehat{12}}}{\l_{\D_0}^{12}} \ ,  \la{oc}
\ee
 where hatted quantities are some particular choice of external dimensions $\widehat{1},\widehat{2}$.
Thus we see that bulk locality places an infinite set of constraints on the boundary OPE coefficients: starting from knowledge of $\l_{\D_n}^{\widehat{12}}$, one can determine $\l_{\D_n}^{ 1  2}$ for any choice of external dimensions $1,2$ (up to an overall factor of $\l_{\D_0}^{12}$).

Let us demonstrate this method for the bulk field $\Phi^2$ in GFF theory (in which case $\b=\D_\p$) with some simple examples of boundary operators.
First, let us consider the form factor $\langle \Phi^2 \,  \mathcal \phi^n \, \mathcal \phi^{n+2}\rangle$. 
Taking the reference dimensions to be, e.g., $\widehat \D_1=\widehat\D_2=\D_\p$, the OPE relations \rf{oc} tell us that:\footnote{We recall that $(\phi^k)_m$ denotes a primary operator in GFF theory  of the schematic form $\phi \Box^{2m} \phi^{k-1}$ and scaling dimension $k\Df+2m$. If $k>2$ and $m>0$ then there are multiple such primaries.}
\be
\l_{(\phi^2)_m}^{\phi^n\, \phi^{n+2}}=0 \ , \qquad m>0\ .
\ee
This is easy to verify since, in free theory, we have a factorization  $\l_{(\phi^2)_m}^{\phi^n\, \phi^{n+2}}\propto \langle (\phi^2)_m \, \phi^2\rangle \langle \phi^n \, \phi^n\rangle $, and the right-hand-side now vanishes for $m>0$ since two-point functions are diagonal.
This implies that the form factor contains only a single boundary block $G_{2\D_\p}$,
which may indeed be seen from its factorization as
$ \langle \Phi^2 \, \phi^n \, \phi^{n+2}\rangle\propto \langle \Phi^2 \phi^2 \rangle\langle \phi^n \phi^n\rangle$.
This is thus an example of the special case considered in Section \ref{spec} containing only the block $\D= |\D_{12}|$ (equalling $2\D_\p$ in this case).

Next let us consider the form factor $\langle \Phi^2 \, \mathcal \phi^n \, \mathcal \phi^n\rangle$. A direct computation in AdS gives 
\be
F(z) =K \, z^{\D_\p} \ , \quad \qquad K=\mu_{\phi^2}^{\Phi^2} \lambda^{\phi^n \, \phi^n}_{\phi^2}
\ee
so this form factor does satisfy \rf{aba} with $\b=\Df$ and the constraints \rf{oc} apply. Taking the same reference dimensions, $\widehat \D_1=\widehat\D_2=\D_\p$, the constraints yield a non-trivial relation between OPE coefficients,
\be
\frac{\ \l_{(\phi^2)_m}^{ \phi^n\, \phi^n}\ }{\l_{\phi^2}^{  \phi^n\, \phi^n}} = \frac{\ \l_{(\phi^2)_m}^{  \phi \, \phi}\ }{\l_{\phi^2}^{  \phi\, \phi}}\,.
\ee
Since this is a free theory, we can check explicitly that these are identities are correct, and indeed they are since
\bea
\l_{(\phi^2)_m}^{\phi^n\, \phi^n}&\propto\, \l_{(\phi^2)_m}^{  \phi \, \phi} \langle \phi^{ n-1} \, \phi^{ n-1}\rangle\,\qquad \mbox{for all}\ m\geq 0\,.
\eea
with the proportionality constant dependent only on $n$.

As a final comment, note that the relations above were unable to fix the zeroth coefficients $\lambda^{\phi^n \, \phi^n}_{\phi^2}$. This was because the overall normalisation of form factors is always left unfixed by locality. But, in fact, we would like to point out that this always had to be the case. This is because all our computations above are really statements about correlators involving operators with dimensions $\Df$, $2\Df$, etc.\ and a specific kind of BOE, and thus they are valid for a much wider set of theories. For instance,
the results would be unchanged if we had considered the theory of $N$ decoupled Generalized Free Fields $\phi_i$,  setting $\phi\to \phi_1,$\ $\phi^2=N^{-\frac 12}\sum_{i=1}^N \phi_i \phi_i$. The unfixed coefficient, for example $\lambda^{\phi^2 \, \phi^2}_{\phi^2}$ in the $n=2$ case, depends on $N$. This explains why we were unable to fix it, since none of our assumptions prefer one value of $N$ over another.

\subsection{Modification for general asymptotics}
As we vary the external operators, 
the asymptotic behaviour generally changes and may not be as suppressed as assumed in \rf{aba}, so the functionals $\theta^{12,\b}_n$ may not be applicable.  This just means that one must consider functionals with higher values of $\tilde \a$, whose kernels have extra suppression at large $z$. Specifically, assuming 
\be
F(z)\underset{z\to\infty}{\sim} z^{\b + p+1-\e} \ , \quad p \in \mathbb{N}  \ , \qquad \qquad \D_n = 2\b+2n \ , \la{decp}
\ee
we should consider the functionals $\theta_m^{12,\b+p}$.
Being dual to the $n>p$ part of the spectrum, the functionals can bootstrap all of the combinations $\mu_n \, \l_{\D_n}^{12}$ with $n>p$ in terms of the others,
\be
\mu_n^\Psi \, \l_{\D_n}^{12} = - \sum_{m=0}^p \mu_m^\Psi \, \l_{\D_m}^{12}  \, \theta^{12,\b+p}_{n-p} (\D_m)  \ , \qquad n>p\ . \la{bsu}
\ee
Given knowledge of the numbers $c^{12}_m = \mu_m^\Psi \, \l_{\D_m}^{12}$ for $m=1,\ldots, p$ (specifying a choice of solution) and fixing a normalization $c_0^{12} = \mu_0^\Psi \, \l_{\D_0}^{12}$, the following combinations are then independent of $\D_1,\D_2$:
\be
\mu_n^\Psi  =     \frac{1}{\l_{\D_n}^{12}}  C^{12,(p)}_n \ , \qquad  C^{12,(p)}_n := -\sum_{m=0}^p c^{12}_m \,  \theta^{12,\b+p}_{n-p} (\D_m) \ , \qquad \quad n>p \ .
\ee
Eliminating $\mu_n^\Psi$, we obtain a relation between OPE coefficients,
\be
\l_{\D_n}^{12} = \frac{C^{12,(p)}_n }{C^{\widehat{12}, (p)}_n} \  \l_{\D_n}^{\widehat{12}} \ , \qquad n>p \ , \la{ore}
\ee
where $\widehat \D_1,\widehat \D_2$ are some reference dimensions whose corresponding form factor also decays at least as fast as \rf{decp}.

These relations between structure constants are a generalization of the $p=0$ case \rf{oc} above (when $C_n^{12,(0)}=-\mu_0^\Psi \l_{\D_0}^{12} \theta_n^{12,\b}(2\b)$). 
For $p>0$, we have essentially lost $p$ of the functionals due to the slower decay of $F(z)$.\footnote{One way to see that we have `lost' $p$ functionals is that the shifted functionals $\theta_n^{12,\b+p}$ are dual in the sense of \rf{dual} to only the $\D_{n>p}$ part of the spectrum. Another is that they can be obtained from the standard functionals $\theta_n^{12,\b}$ by making $p$ subtractions (see eq.\ \rf{eq:funcrels} and discussion there).}
As a result, in \rf{ore}, we lose $p$ of the relations between structure constants; however, an infinite set of relations still remain. This simply indicates that there are $p+1$ independent solutions with the same spectrum, external dimensions and asymptotics. Indeed, whenever there are multiple solutions, this \textit{must} happen, and the asymptotic of $F(z)$ \textit{must} become correspondingly less suppressed.

One source of multiplicity of solutions is the degeneracy of generic higher-trace states $(\phi^n)_m$ (but note this is somewhat particular to the free theory). However, as we will show, there are multiple solutions even for the non-degenerate state $(\phi^2)_1$. Let us consider in detail the form factor $\langle \Phi^2 \,  \phi^2 \, (\phi^2)_1 \rangle $ in GFF theory. The BOE spectrum is $\D_n=2\b+2n$ with $\b=\Df$. By considering the flat space limit, we estimate the large $z$ asymptotic to be 
 \be
 F(z) \sim z^{\D_\p+1} =  z^{\b+1}\ , \la{esta}
 \ee
 i.e.\ one power of $z$ harder than the contact term $\langle \Phi^2 \,  \phi^2 \, \phi^2 \rangle$, since the operator $(\phi^2)_1$ contains two extra derivatives and $z \sim -s \sim p^2$ in the flat space limit. 
 
 The estimated asymptotic \rf{esta} satisfies \rf{decp} with $p=1$, i.e.\ requiring one subtraction. This means there are \textit{two} independent solutions of the locality constraints with this BOE spectrum, external dimensions, and asymptotic (or softer). A moment's reflection indeed reveals another such solution: 
 instead of the derivatives in the operator $(\phi^2)_1$, one can produce its excess dimension by coupling to a heavier free field $\Phi_2$. The form factor $\langle (\Phi_1 \Phi_2) ( \phi_1\phi_2) (\phi_2^2) \rangle$, in a GFF theory with two scalars of dimensions $\D_1 = \D_\p-1$ and $\D_2=\D_\p+1$, has  the same external dimensions ($2\D_\p$ and $2\D_\p+2$) and BOE spectrum ($2\D_\p+2n$) as above. Being a simple contact without derivatives, we expect this solution to have the more suppressed asymptotic $F(z) \sim z^\Df = z^\b$.

 Let us bootstrap these two solutions using the locality constraints. Starting from `initial data' $c_0$,$c_1$, all the other combinations $c_n = \mu_n^\Psi \,  \l_{\D_n}^{12}$ are bootstrapped by the sum rules \rf{bsu},
\begin{align}
&F(z)  = \sum_{n=0}^\infty 
c_n \, G_{2\D_\p+2n}^{2\D_\p,2\D_\p+2n} (z) \ , \la{BFE} \\
&c_n =  \frac{(-1)^{n-1} \Gamma(-1 + \D_\p + n) \Gamma(
  1 + \D_\p + n)}{\Gamma(-\tfrac{d}{2} + 2 \D_\p + 2 n)}   \la{cne} \\
  &\quad \times 
  \Big[    \frac{
   4  (n-1) \Gamma(2-\tfrac{d}{2} + 2 \D_\p + n)}{(d-2 - 4 \D_\p) (d - 4 \D_\p - 2 n) \Gamma(\D_\p-1) \Gamma(
     \D_\p+1) \, n!} \, c_0  + \frac{ \Gamma(1-\tfrac{d}{2} + 2 \D_\p + n)}{
   \Gamma(\D_\p) \Gamma(\D_\p+2) \,  (n-1)!} \, c_1 
\Big] \no \ .
\end{align}
One coefficient $c_0$ may be seen as a normalization ambiguity, while the other $c_1$ corresponds to the choice of theory.

Let us identify the values of $c_1$ corresponding to our two solutions of interest by explicitly computing the form factors. A bulk computation of the GFF form factor $\langle \Phi^2 \, \phi^2 \, (\phi^2)_1 \rangle$ gives 
\be
F(z) = K \, z^{\D_\p} \left[\frac{2(\D_\p+1)}{d-2-2\Df} z +1\right] \ ,\qquad K=\frac{\lambda^{\phi^2\, (\phi^2)_1}_{\phi^2}}{\lambda^{\phi \, \phi}_{\phi^2}}
\la{s1}
\ee
where we fixed the normalisation by setting $\mu_{\phi^2}^{\Phi^2}=1/\lambda^{\phi \,  \phi}_{\phi^2}$. This result is consistent
with the bootstrap result \rf{cne} upon expanding it in blocks as \rf{BFE} and making the choice
\be
c_0 = K  \ , \qquad c_1 = K \, \frac{2 \D_\p(1 + \D_\p) (d-4-2\D_\p)}{(d-2-2\Df)(d-2 - 4 \D_\p)} \ . \la{ini}
\ee
With these `initial data', 
 the structure constants are bootstrapped by \rf{ore} in terms of some `reference' ones, which we take to be $\l_{(\phi^2)_n}^{\phi\,  \phi}$,
\begin{align}
\frac{\ \l_{(\phi^2)_n}^{\phi^2\, (\phi^2)_1}\ }{\l_{\phi^2}^{\phi^2\, (\phi^2)_1}} &= 
 \frac{(\Df+n)(d-2-2\Df-2n)}{\Df(d-2-2\Df)}
 \, \frac{\ \l_{(\phi^2)_n}^{\phi \, \phi}\ }{\l_{\phi^2}^{\phi \, \phi}} \ .
\end{align}
Using the known formula for the reference structure constants $\l_{(\phi^2)_n}^{\phi \, \phi}$  in GFF theory~\cite{Fitzpatrick:2011dm}, we obtain
\be
\left(\l_{(\phi^2)_n}^{\phi^2\, (\phi^2)_1}\right)^2 =  \frac{2K^2\, (\D_\p+n)^2(d-2\Df-2n-2)^2 (\Df)_n{}^2(\Df+1-\tfrac{d}{2})_n{}^2}{(d-2-2\Df)^2\D_\p^2 \, n!\, (\tfrac{d}{2})_n (2\Df+n+1-d)_n (2\Df+n-\tfrac{d}{2})_n} \ .  \la{struc1}
\ee

Now let us turn to the other solution $\langle (\Phi_1 \Phi_2) ( \phi_1\phi_2) (\phi_2^2) \rangle$ described above. In free theory, this form factor factorizes as $\langle (\Phi_1 \Phi_2) ( \phi_1\phi_2) (\phi_2^2) \rangle \propto \langle (\Phi_1 \Phi_2)\, \phi_1 \, \phi_2 \rangle \langle \Phi_2 \, \phi_2 \rangle$ and, as a result, the form factor is proportional to a simple contact (cf.\ the form factor \rf{F0}),
\be
F(z) = K z^{\D_\p}  \ , \qquad \qquad K = \frac{ \ \l^{(\phi_1\phi_2) \, (\phi_2^2)}_{(\phi_1\phi_2)} \ }{\l^{\phi_1\, \phi_2}_{(\phi_1\phi_2)}} \la{s2} \ ,
\ee
with the normalisation fixed as $\mu^{(\Phi_1\Phi_2)}_{(\phi_1\phi_2)} = 1/\l^{\phi_1\, \phi_2}_{(\phi_1\phi_2)}$. Expanding  \rf{s2} in blocks, it matches the bootstrap result \rf{cne} with\footnote{This solution can also be identified by its more suppressed asymptotic, which implies it can be bootstrapped directly using \rf{oc} without `subtractions', in terms of only one normalization constant $c_0$.}
\be
c_0 =K \ , \qquad \quad c_1 = K\,  \frac{2(\Delta_\phi^2-1)}{d-2-4\Df} \ .
\ee
To bootstrap the structure constants, we will now take as our `reference' $\lambda_{(\phi_1\phi_2)_n}^{\phi_1 \, \phi_2}$. Then \rf{ore} gives the identity 
\be
\frac{\lambda_{(\phi_1\phi_2)_n}^{(\phi_1 \phi_2)\, (\phi_2^2)}}{ \ \l^{(\phi_1\phi_2) \, (\phi_2^2)}_{(\phi_1\phi_2)} \ } = \frac{ \ \lambda_{(\phi_1\phi_2)_n}^{\phi_1\,  \phi_2} \ }{\l^{\phi_1\, \phi_2}_{(\phi_1\phi_2)} }
\ee
The `reference' structure constants on the right-hand side are well known \cite{Fitzpatrick:2011dm}, giving 
\be
\left( \lambda_{(\phi_1\phi_2)_n}^{(\phi_1 \phi_2)\, (\phi_2^2)} \right ) ^2 = \frac{2K^2\,  (\D_1)_n\, (\D_2)_n \, (\D_1+1-\tfrac{d}{2})_n{}\,(\D_2+1-\tfrac{d}{2})_n{}}{ n!\, (\tfrac{d}{2})_n (\D_1+\D_2+n+1-d)_n (\D_1+\D_2+n-\tfrac{d}{2})_n} \ . \la{rla}
\ee
It is relatively difficult to obtain \rf{struc1} and \rf{rla} directly from CFT four-point functions in general spacetime dimension --- which speaks to the power of the present approach --- but, to the extent that we were able to do this, we found perfect agreement with our results.

\section{Form factors in the flat space limit}
\label{sec:flat}
\subsection{Taking the flat space limit}
In this section, we will study the flat space limit of AdS form factors, mimicking the logic for boundary correlators \cite{QFTinAdS,Komatsu:2020sag,Cordova:2022pbl,vanRees:2022itk}. Concretely, we consider a gapped QFT in AdS. This induces a {\em family} of boundary CFTs, labeled by the mass gap in units of the inverse AdS radius. The flat space limit involves two steps. Firstly, we must take large AdS radius $R$, while keeping physical masses $m\sim \Delta/R$ fixed. Since the theory is gapped, it follows that all scaling dimensions become very large in this limit. We will therefore use the terms `large $R$' and  `large $\Delta$' interchangeably (and in fact `large $\Delta$' is the most natural formulation from the CFT perspective).
Secondly, we must perform a suitable analytic continuation into appropriate scattering kinematics. In the process, there are several subtleties to worry about, like commuting limits, whether quantities blow up or stay finite, and so on --- but here we will mostly ignore these subtleties. Technically our arguments are very similar to those of \cite{Cordova:2022pbl}, but we will not push our analysis as far as in that reference.

Let us begin by motivating the analytic continuation and the mapping of kinematics. In global AdS coordinates we have,
\ba
\langle \Psi(X) \mathcal O(P)\rangle\sim \frac{1}{(-2 P\cdot X)^{\Delta}}\quad \underset{R \to \infty}= \quad 2^{-\Delta}\,e^{m n_P\cdot x}\,, \qquad x^\mu\equiv r\, n_X^\mu \ .
\ea
We would like this to become a one-point form factor of the bulk field $\psi$, i.e.\ a plane wave. This can be achieved by the continuation
\ba
n_P^0\to - k^0/m\,, \qquad n_P^i=i k^i/m\,, \qquad x^0\to i x^0 \ ,
\ea
with $k^\mu$ Lorentzian and $k^2=-m^2$. The expectation is that, for a generic correlator involving bulk and boundary fields, in the flat space limit this continuation maps local boundary operator insertions to in/out states of definite on-shell momenta~\cite{Hijano:2019qmi}:
\begin{multline}
2^{\sum_i^m\Delta_i}\langle 0|\Psi_1(X_1) \ldots \Psi_n(X_n) \mathcal O_1(P_1)\ldots \cO_m(P_n)|0\rangle_R\\ \underset{R\to \infty}\to \langle 0|\Psi(x_1)\ldots \Psi_n(x_n)|k_1\ldots k_n\rangle \ ,
\end{multline}
where we explicitly indicate dependence on the AdS radius $R$. 

The perturbative validity of the above follows from a straightforward adaptation of the arguments of \cite{Komatsu:2020sag}. A non-perturbative proof for an arbitrary number of insertions would be significantly more difficult. Here we will restrict our attention to the single bulk insertion, two-point AdS form factors that we've been studying so far. In this case we have
\ba
2^{\Delta_1+\Delta_2}\langle \Psi(X) \mathcal O_{\Delta_1} (P_1) \mathcal O_{\Delta_2}(P_2)\rangle_{R} &= \frac{z^{-\frac{\D_1+\D_2}2}F_{R}(z)}{(-P_1\cdot X)^{\Delta_1}(-P_2\cdot X)^{\Delta_2}}\\
&\underset{R\to \infty}{=}  e^{i(k_1+k_2)\cdot x}\,\mathcal F(s) =:\langle 0|\Psi(x)|k_1,k_2\rangle \ ,
\ea
which therefore sets
\ba
\boxed{
\mathcal F(s)= \lim_{R\to \infty} z_s^{-\frac{\D_1+\D_2}2}F_{R}(z_s)\,. 
}
\label{eq:flatF}
\ea
In this formula the cross-ratio is written
\ba
z=\frac{-2 R^2 P_{12}}{(-2 P_1\cdot X)(-2 P_2\cdot X)} \underset{R\to \infty}{\to} z_s\equiv  -\frac{s-(m_1+m_2)^2}{4 m_1 m_2}\label{eq:zmapp}
\ea
with
\ba
s\equiv -(k_1+k_2)^2\,, \qquad m_{1,2}=\frac{\Delta_{1,2}}{R}\,.
\ea
The regime of physical kinematics is then $s>(m_1+m_2)^2$, i.e.\ negative $z$ with a small negative imaginary part.

From the above, it seems that the AdS form factor directly becomes the flat space one. In fact, we do not expect \reef{eq:flatF} to literally hold for arbitrary kinematics: what is meant by that equation is that we should first take the limit in a region where it exists, and then analytically continue. In the process of taking the analytic continuation, we may encounter singularities that were not present in $F_R(z)$ for any fixed $R$ and $z$, since the limits do not commute in general. The logic is entirely analogous to the analysis of boundary correlators in \cite{Cordova:2022pbl} and could presumably be repeated straightforwardly in the present context.\footnote{Unlike there however, the lack of positivity in the BOE would surely lead to less powerful results.}

Before we proceed, let us comment on what is expected of a form factor in flat space, and how our proposal above is consistent with these expectations. A basic feature is that, for physical kinematics $s>(m_1+m_2)^2$, there should be a branch cut corresponding to exchange of double or multiparticle states. This translates into the cut of the AdS form factor with $z<0$. More importantly, in this context, the meaning of locality is that, as we vary the Mandelstam invariant $s$, the only singularities we encounter on the complex plane should have a definite physical meaning, such as poles associated to stable particles, and branch-cuts for multiparticle exchange. As we mentioned above, such singularities can in principle arise in the flat space limit. However, if we assume that locality of the AdS form factor (i.e.\ analyticity for $z\geq 1$) survives the flat space limit, it would then imply that the momentum space form factor is 
% local 
analytic for $s\leq (m_1-m_2)^2$. This makes sense: the presence of a pole at or below this critical value would mean that we would not be considering form factors of stable particles, so that the premise of our whole analysis would be incorrect.

Below we will see, both in specific examples and more generally under reasonable assumptions, that the flat space form factors we obtain do indeed have the good analyticity properties set out above, these being essentially inherited from their AdS counterparts.

\subsection{A phase shift formula}
We will now derive an expression for the flat space form factor in physical kinematics in terms of the boundary CFT data. The idea is simple: we will use the formula \reef{eq:flatF} directly for these kinematics $s>(m_1+m_2)^2$ (or equivalently $z<0$ with a small negative imaginary part), while simultaneously decomposing the form factor using the BOE. We will have to assume that the flat space limit commutes with the limit of physical kinematics.
This seems reasonable given that it is true in the four-point function context \cite{Cordova:2022pbl} --- here we will see \textit{a posteriori} that this assumption is justified in a number of examples. 

Our goal is then to compute:
\ba
\lim_{R\to \infty} z^{-\frac{\D_1+\D_2}2} F_R(z_s)=\sum_{\Delta}\lim_{R\to \infty} c_{\Delta} z^{-\frac{\D_1+\D_2}2}G_{\Delta}^{12}(z)\ , \qquad z<0\ .
\ea
We will need expressions for boundary blocks in the flat space limit and on-shell kinematics. Here we will set $\Delta_1=\Delta_2\equiv\Df$, leaving the more general but cumbersome formulae to Appendix \ref{app:flat}. A short calculation gives:
\ba
z^{-\Df}G_{\Delta}(z)&\underset{z<0}=\frac{e^{-i\frac{\pi}2(\Delta-2\Df)}}{(1-z)^{\Df}} \left(\frac{z}{z-1}\right)^{\frac \Delta2-\Df}\, _2F_1\left(\frac{\Delta}2,\frac{\Delta-d+2}2,\Delta-\frac{d-2}2,\frac{z}{z-1}\right)\\
&\hspace{-12pt}\underset{\Delta,\Df\to \infty}{\sim}
\frac{e^{-i\frac{\pi}2(\Delta-2\Df)}}{(-z)^{\Df}}  \frac{\left[-4\rho^2(\mbox{$\frac{z}{z-1}$})\right]^{\Delta/2}}{\sqrt{1-\rho^4(\mbox{$\frac{z}{z-1}$})}\,\left[1+\rho^2(\mbox{$\frac{z}{z-1}$})]\right]^{\frac{d-2}2}}\\
&\, \sim 2\frac{e^{-i\frac{\pi}2(\Delta-2\Df)}}{\hat c_{\Delta}^{\mbox{\tiny free}}}\, \mathcal N_{\Df}(\Delta,4(1-z))\,,\qquad \Delta>2\Df\ ,
\ea
where the last line is valid up to exponentially suppressed corrections. The function $\mathcal N_{\Df}(\Delta,s)$ is a unit normalized gaussian in the variable $\Delta$, of variance $O(\Df)$ and centered around~$(\Delta/\Df)^2=s$:
\ba
\mathcal N_{\Df}(\Delta,s)=\sqrt{\frac{1}{\pi \Df(s-4)}}\, \exp\left[-\frac{(\Delta-\sqrt{s}\Df)^2}{\Df(s-4)}\right]\,.
\ea
A similar calculation shows that
blocks with $\Delta<2\Df$ are exponentially suppressed.\footnote{More properly, such blocks are exponentially suppressed for sufficiently negative $z$. We then {\em define} the analytic continuation of the form factor for all negative $z$ by explicitly subtracting out those blocks.}
As for the quantity $\hat c_{\Delta}^{\mbox{\tiny free}}$, it is related to a contact form factor:\footnote{This form factor and its expansion can be derived in a similar fashion to what was done in Section \ref{Scont}.} 
\ba
F^{(\Phi_1\Phi_2)}_{\phi_1\, \phi_2}(z)=z^{\frac{\D_1+\D_2}2}=\sum_{n=0}^\infty (-1)^n\,  \hat c_{\Delta_n}^{\mbox{\tiny free}} \, G_{\D_1+\D_2+2n}^{12}(z)
\ea
with
\ba
\hat c_{\Delta}^{\mbox{\tiny free}}=\frac{\Gamma \left(\frac{ \Delta +\Delta _{12}}2\right) \Gamma \left(\frac{\Delta
   -\Delta _{12}}2\right) \Gamma \left(\frac{\Delta-d +\Delta _1+\Delta
   _2}2\right)}{\Gamma \left(\Delta _1\right)\Gamma \left(\Delta _2\right)  \Gamma \left(\frac{ \Delta -\Delta _1-\Delta
   _2+2}2\right) \Gamma \left(\Delta -\frac{d}{2}\right)} \ .
\ea
Plugging the expressions for the block  into the BOE (omitting those with $\Delta<2\Df$), we obtain the following {\em phase shift formula} for the on-shell form factor:
\ba
\boxed{\mathcal F(s)= \lim_{R\to \infty} \sum_{\Delta>\D_1+\D_2} 2\, e^{-i\frac{\pi}2(\Delta-\D_1-\D_2)} \left(\frac{c_\Delta}{\hat c_{\Delta}^{\mbox{\tiny free}}}\right)\, \mathcal N_{\D_1,\D_2}(\Delta,s)\ ,
}\label{eq:phaseshift}
\ea
which we presented here in the general case $\D_1\neq \D_2$. The function $\mathcal N_{\D_1,\D_2}$ is defined in Appendix \ref{app:flat}. It remains a gaussian but now localizes at $s= 4\left(\frac{\Delta}{\D_1+\D_2}\right)^2$. The formula above is only valid for physical kinematics $s>(m_1+m_2)^2$ with a small positive imaginary part. 

\subsection{Applications}

Let us now look at some examples of the formulae above. Firstly, consider the
form factors $F^{(k)}$ of operators of the schematic form $\partial^k\Phi\partial^k \Phi$ in free theory, which we determined in Section \ref{Scont} to take the form:
\ba
F^{(k)}(z)=z^{\Df+k}\ .
\ea
Using \reef{eq:flatF} and setting $m_1=m_2=1$, we immediately get
\ba
 \mathcal F^{(k)}(s)\propto (s-4)^k \label{eq:Fk}
\ea
which is the correct result corresponding to the flat space form factor $\langle 0|\partial^k\Phi\partial^k \Phi | k_1, k_2\rangle$. Alternatively we can use the phase shift formula. The BOE data is now
\ba
c_{\Delta_n}^{(k)}\underset{\Df,n\to \infty}=(s_n-4)^{k} (-1)^n \hat c^{\mbox{\tiny free}}_{\Delta_n}\,, \qquad s_n\equiv \left(\frac{\Delta_n}{\Df}\right)^2 \ .
\ea
Plugging this into \reef{eq:phaseshift}, a simple computation leads again to \reef{eq:Fk}.

As a different example, we can compute the flat space limit of a local block with $\Delta=\Delta_b<2\Df$ corresponding to the exchange of a bound state of mass $m_b=\Delta_b/R$. We will obtain a finite answer if we multiply the local block with the correct prefactor. Let us set
\ba
\tilde c_{\Delta_b}^{\mbox{\tiny free}}= \frac{\hat c_{\Delta_b}^{\mbox{\tiny free}}}{\sin\left[\frac{\pi}2 (\Delta-2\Df)\right]}
\ea
We will compute
\ba
\mathcal F_b(s)& =g\lim_{\Delta_b,\Df\to \infty} (\Delta_b \tilde c_{\Delta_b})  \left[z^{-\Df} \mathcal L^{\Df-1}_{\Delta_b}(z_s)\right]
\ea
The justification for this specific prefactor is that it corresponds to the one obtained by the AdS Lagrangian computation of Section \ref{localS} since, in the notation of that section, we have $\Delta_b \tilde c_{\Delta_b}\propto a^{12}(\Delta_b)$. In other words, this is the prefactor we must include to obtain a good flat space limit, where $g$ remains constant and finite in units of the mass gap to some power. Using the phase shift formula we get
\ba
\mathcal F_b(s)
&=-g \Delta_b \sum_{n=0}^\infty e^{-i \frac{\pi}2(\Delta_n-2\Df)}\left(\frac{\tilde c_{\Delta_b}^{\mbox{\tiny free}}}{\hat c_{\Delta_n}^{\mbox{\tiny free}}}\right)\theta_n^{\Df-1}(\Delta_b) \mathcal N_{\Df}(\Delta_n,s)\\
&\sim -\frac {g \Delta_b} 2 \left(\frac{\tilde c_{\Delta_b}^{\mbox{\tiny free}}}{\hat c_{\Delta_n}^{\mbox{\tiny free}}}\right)\, \theta_{n_s}^{\Df-1}(\Delta_b)\ , \qquad \left(\frac{\Delta_{n_s}}{\Df}\right)^2=s \ . \\
&=g\,\frac{2m_b^2}{\pi} \frac{1}{m_b^2-s}
\ea
We see that the form factor takes precisely the expected form in the flat space limit. 

We can combine the phase shift formula with this result to get a dispersion relation for the form factor. For simplicity we will do the derivation for $\af=\Df=\D_1=\D_2$. We start from \reef{eq:dispF} and take the flat space limit:
\begin{align}
\no &F(w)=\sum_{0\leq \Delta\leq 2 \Df} c_{\Delta}\mathcal L_{\Delta}^{\Df}(w)+\int_{-\infty}^0 \frac{\ud z}{\pi}\,\frac {w^{\Df+1}}{z(z-w)}\mathcal I_z\left[z^{-\Df}\left(F(z)-\sum_{0\leq \Delta\leq 2\Df} c_{\Delta} G_{\Delta}(z)\right)\right]\\
\Rightarrow \quad & \mathcal F(s)=\sum_{i} g_i \frac{s-4}{(m_i^2-4)(m_i^2-s)}+\int_{4}^\infty \frac{\ud s'}{\pi} \frac{4-s}{(4-s')(s-s')} \, \nu({s'})
\end{align}
where, using the phase shift formula for the integrand,
\ba
\nu(s)\equiv \mathcal I_{s} \mathcal F({s})=\sum_{\Delta>\D_1+\D_2} 2\, \sin\left[\frac{\pi}2(\Delta-2\Df)\right] \left(\frac{c_\Delta}{\hat c_{\Delta}^{\mbox{\tiny free}}}\right)\, \mathcal N_{\Df}(\Delta,s)\,, 
\ea
Note that the sum over states localizes in a small window around $(\Delta/\Df)^2-s=O(\sqrt{\Df})$. The above establishes an analytic continuation of the flat space form factor from physical to all complex kinematics. In particular, assuming the expression makes sense  (i.e.\ the integral converges), the above demonstrates that the flat space form factor is indeed local in the sense discussed previously. 

\subsection{Extremality and integrability}
To conclude this section, let us make a connection to the theory of 2d integrable models. In such models, form factors can be constructed appealing to two conditions known as Watson's equations (see e.g. \cite{Mussardo:2020rxh}). Let us focus on the two-point form factors considered in this work. The first condition is that form factors are local, i.e.
\ba
\mathcal F(e^{2\pi i} s)=\mathcal F(s) \ .
\ea
This means that form factors do not have branch cuts along the negative $s$ axis. Using the mapping $z=1-s/4$ given in \rf{eq:zmapp}, this translates to absence of a discontinuity for $z>1$, i.e.\ the locality conditions we have discussed in this work. The second condition, which holds specifically for 2d integrable models, is that the discontinuity of the form factor for physical kinematics should be given by the S-matrix. This translates to:
\ba
\mathcal F(s+i\epsilon)=\mathcal F(s-i\epsilon) \mathcal S(s+i\epsilon)\ , \qquad s>4 \ , \label{eq:watson}
\ea
where $\mathcal S(s)$ is the two-to-two S-matrix. We will now see that this condition can be derived from our construction assuming that the BOE data is of a special kind in the flat space limit. In fact, our argument actually shows the above holds in general spacetime dimension upon replacing $\mathcal S(s)\to \mathcal S_{\ell=0}(s)$ (the spin-0 phase shift). The key assumption is:
\begin{itemize}
\item In the flat space limit $\Df\to \infty$, the BOE is effectively captured by a single analytic family of states of dimensions $\Delta_n=2\Df+2n+\gamma_n$, with $\gamma_n$ slowly varying, i.e. $\partial_n\gamma_n=O(1/\Df)$.
\end{itemize}
Physically this means that the BOE spectrum (in the scalar channel) has no `particle production' in the flat space limit, so that only double-trace type operators contribute.  Under this assumption, the phase shift formula simplifies to
\ba
\mathcal F(s)=\sum_{\Delta>2\Df} 2\, e^{-i\frac{\pi}2(\Delta-2\Df)} \left(\frac{c_\Delta}{\hat c_{\Delta}^{\mbox{\tiny free}}}\right)\, \mathcal N_{\Df}(\Delta,s)=e^{-\frac{i \pi}2\gamma(s)} g(s)\,, 
\ea
with 
\ba
g(s)&=\sum_{n=0}^\infty 2\,(-1)^n\left[\left(\frac{c_\Delta}{\hat c_{\Delta}^{\mbox{\tiny free}}}\right)\, \mathcal N_{\Df}(\Delta,s)\right]\bigg|_{\Delta=2\Df+2n+\gamma_n}\,, \\
\gamma(s)&= \gamma_n \bigg|_{n=\lfloor \Df\frac{\sqrt{s}-2}2\rfloor}
\ea
and the flat space limit implied everywhere. In previous work \cite{Cordova:2022pbl} it was shown that $e^{-i\pi \gamma(s)}$ defined above is precisely the 2-to-2 S-matrix of the integrable model, $e^{\mp i\gamma(s)}=\mathcal S(s\pm i\epsilon)$ (or the spin-0 phase shift in general dimension), from which now follows \reef{eq:watson}.

\section{Discussion} \la{Disc}
We have initiated a systematic study of bulk reconstruction from an operator perspective. Specifically, we reformulated it 
as the problem of finding solutions to 
a set of functional sum rules on the BOE and boundary OPE data. They arise by imposing locality of correlators involving one bulk and two boundary insertions --- `AdS form factors'.  A nice feature of these sum rules is that they can be chosen to be dual to Generalized Free Field solutions in the sense of eq.\ \rf{dual}. This means that they are especially well adapted for studying QFTs in AdS that are perturbatively close to a free theory. Thus, as promised in the introduction, this formalism allows one to canonically construct local bulk fields in large $N$ CFTs.

Let us 
discuss how our construction must be modified to account for bulk theories with local symmetries, leaving a detailed account for future work \cite{part3}. For the purposes of discussion, let us concentrate on internal, local symmetries.
In this context, consider what would happen if we try to reconstruct a local bulk operator that is charged under the gauge group. 
By gauge invariance, we know that the operator must get dressed with gauge flux, perhaps in the form of a Wilson line. The flux stretches out to the AdS boundary and should show up as a breakdown of locality.\footnote{An interesting alternative is that this flux might instead get attached onto features of the background geometry, or in CFT language, onto specific features of a non-vacuum state \cite{Bahiru:2022oas,Bahiru:2023zlc}. This should lead to a breakdown of locality that is exponentially suppressed, rather than power law as discussed here.}
Concretely, form factors involving the boundary conserved current dual to the bulk gauge field should be sensitive to this dressing as a failure of conservation in the following sense. Supposing we attach a Wilson line to the bulk operator, reaching the boundary at some point $x_0$, let us schematically denote the full object as $\Phi_{x_0}(x,u)$. Then current conservation gives:
\be \begin{split}
\frac{\partial}{\partial x_1^\mu}\, \langle  J^\mu(x_1) \bar{\mathcal \phi}(x_2) \Phi_{x_0}(x,u)\rangle&= i  q \, \delta^{(d)}(x_1-x_0) \, \langle \bar{\mathcal  \phi}(x_2) \Phi_{x_0}(x,u)\rangle\\
&\qquad \qquad - i  q \, \delta^{(d)}(x_1-x_2) \, \langle \bar{\mathcal  \phi}(x_2) \Phi_{x_0}(x,u)\rangle
\end{split} \ee
with $q$ the charge of $\phi(x)$. For instance, if we choose $x_0=x$ so that the Wilson line lies along the orthogonal projection of $\Phi$ to the boundary, the  delta function takes the form $\delta(z-1)$. Thus we expect the AdS form factor to have a singularity, and for locality to break down at $z=1$. More generally, dressed operators will lead to failures of conservation of $J^\mu$ and the breakdown of locality away from boundary operator insertions. The good news is that the above suggests this breakdown should be controlled by the Ward identity plus a choice of dressing. Thus our sum rules should still be applicable, albeit now with a non-trivial (and hopefully controllable, at least for large $N$) right-hand side, 
\ba
\sum_\Delta c_{\Delta} \theta_n(\Delta)=(\mbox{fixed by conservation + dressing})\,.
\ea
Similar comments generalize straightforwardly to dressings of operators in the presence of gravity, replacing the conserved current with the stress tensor. We hope to pursue further investigations along these lines in the near future.

In spite of the above discussion, we would like to emphasize that the perspective pursued in this work, namely of non-gravitational QFTs in AdS space, is not without its uses. Indeed, the study of QFTs in AdS has received increasing interest in recent years \cite{QFTinAdS,Cordova:2022pbl,Komatsu:2020sag,Behan:2021tcn,vanRees:2022itk,Antunes:2021abs}, as this setup allows QFT questions to be formulated in terms of CFT ones, which can then be readily adressed with rigorous bootstrap techniques. The present work can be seen as a way to systematically include additional constraints on a CFT, other than crossing and unitarity, for it to be compatible with the existence of a local AdS bulk description. In this paper, we have begun to explore this analytically. For example, in the context of GFF theory, we showed explicitly that bulk locality imposes infinitely many relations between OPE coefficients. Similarly, interesting CFTs without a stress-tensor, such as long-range models, can be described by QFTs in AdS and the kind of sum rules described in this work should then be very useful in bootstrap applications to impose extra constraints on the OPE data. In fact, the simplest such constraints, associated with the existence of a bulk free field\footnote{As we saw in Section \ref{spec}, free fields in the bulk are a special case where the BOE contains only two blocks.}, have already been put to use in such applications~\cite{Behan:2018hfx,Behan:2020nsf,Behan:2021tcn}. Our work adds significantly new constraints to account for the existence of composite bulk fields. The results may be seen as an AdS generalization of the kind of logic that was previously used in bootstrap studies of S-matrices and form factors \cite{Karateev:2019ymz,Karateev:2022jdb}.

One of the outcomes of the present work is that flat space form factors can be determined from the boundary CFT. Here we have focused on the case where the bulk theory is gapped, but it would be interesting to pursue a more detailed understanding of the dictionary for gapless theories. This would then allow us to explore situations where the bulk theory suffers from infrared divergences and there are difficulties in defining a finite S-matrix. AdS/CFT suggests a natural regularization for these divergences and provides non-perturbative, well-defined observables --- boundary CFT correlators --- which can perhaps be used to define good S-matrices in the bulk.\footnote{As well as the well-known infrared divergences in 4d (see \cite{Prabhu:2022zcr} for a recent discussion), infrared issues can also arise in integrable 2d models (see e.g. \cite{HLT}). Boundary correlators in AdS have indeed been considered as an infrared regulator in that context \cite{Beccaria:2020qtk}.} It seems to us that, to make progress in this direction, it would be important to understand how LSZ reduction of bulk correlators in flat space is related to the push map in AdS/CFT --- something for which the constructions presented in this paper should be very useful. 

Although our language has been that of QFTs in AdS space, our results also have immediate applications to boundary conformal field theories (BCFTs). Indeed, that merely corresponds to the case where the QFT happens to be a conformally invariant CFT$_{d+1}$. In that case the bulk fields can be labelled by representations of the (bulk) conformal group $SO(d+2,1)$ and, in particular, by bulk scaling dimensions $\tilde \Delta^{\Fi}$. In these circumstances, one can make a Weyl mapping from AdS to a flat metric, which we can think of as a half-space, or equivalently the interior of the $d+1$ dimensional ball. The mapping of correlators is simply given by
\ba
(-X\cdot I)^{\tilde \Delta}\Fi(X)\bigg|_{\mbox{\tiny AdS}}=\Fi(X)\bigg|_{\mbox{\tiny flat}} \ .
\ea
In particular, this means BCFT correlators are trivially related to the ones considered in this work, so our results apply equally well to that case.  In BCFT there is, of course, also an OPE available between bulk fields, and this fact has been exploited in the past to obtain constraints on both BOE and bulk OPE CFT data \cite{Liendo:2012hy} by bootstrapping bulk two point functions. It would be very interesting to combine these with the set of bulk locality sum rules that we developed in the present work, and to see what improvements they may add.

As a final note, let us point out that, on a technical level, our construction is closely analogous to what was done for the crossing equation in CFTs \cite{Mazac:2018ycv}, especially in 1d where there is a single cross-ratio, as for our locality equation. However, our construction of functional kernels was significantly easier for two reasons.  Firstly, it was sufficient for us to focus on functionals whose actions have simple zeros. While some functionals with double zeros relevant to the present setup follow from those presented in \cite{Kaviraj:2018tfd}, initial explorations did not seem to show them as immediately useful. Secondly, the functionals here act directly on blocks rather than differences of direct and cross channel blocks. Since boundary blocks are eigenfunctions of a differential operator, one may naturally wonder if our constructions can be interpreted 
as obtaining a `basis' for some space. In follow-up work \cite{part2}, we will make this idea precise. 
We will demonstrate that the functionals constructed here indeed form a basis in a well-defined sense.
We will also show that it is possible to characterize and construct a broad class of many other bases, which are dual to BOEs with non-trivial spectra. An optimistic hope would be to generalize this line of thought to the crossing equation and eventually obtain exact `extremal' interacting solutions to the crossing equation, at least for 1d CFTs.

\subsection*{Acknowledgements}
The authors are grateful to Y.~He, J.~Henriksson, E.~Lauria and M.~Meineri for interesting discussions.
This work was co-funded by the European Union (ERC, FUNBOOTS, project number 101043588). Views and opinions expressed are however those of the author(s) only and do not necessarily reflect those of the European Union or the European Research Council. Neither the European Union nor the granting authority can be held responsible for them.
NL was supported by the Institut Philippe Meyer at the {\'E}cole Normale Sup{\'e}rieure in Paris.

\newpage

	\appendix	

\section{Integral identities for exchange diagram \la{inti}}
In this appendix to Section \ref{localS}, we give some integral identities that were used to write the exchange diagram \rf{ES} in the form \rf{lb}. The integral over $X'$ in \rf{ES} can be done using the following identity 
\ba
\int \ud X' \, \frac{1}{\prod_{i=1}^3 (-2P_i\cdot X')^{\Delta_i}}=\frac{\pi^{\frac{d}2}}{2} \frac{\Gamma\left(\frac{\sum_i \Delta_i-d}2\right) \prod_{i<j}^3 \Gamma(\delta_{ij})}{\prod_{i=1}^3 \Gamma(\Delta_i)} \frac{1}{\prod_{i<j}^3 (-2P_{i}\cdot P_j)^{\delta_{ij}}} \ ,
\ea
where $\delta_{12}:=\frac{\Delta_1+\Delta_2-\Delta_3}2$ and so on. Then the integral over $Q$ is done using the identity
\ba
\int_{\mathbb R^d} \ud Q \, \prod_{i=1}^3\frac{\Gamma(\Delta_i)}{(-2P_i\cdot Q)^{\Delta_i}}= 2\pi^{\frac d2} \int_0^\infty \prod_{i=1}^3 \frac{\ud t_i}{t_i} t_i^{\Delta_i} \exp\left[\left(\sum_{i=1}^3 t_i P_i\right)^2\right]\,,\quad  \quad \sum_{i=1}^3 \Delta_i=d \ .
\ea
In the particular case $P_3\equiv X$, i.e. $P_3^2=-1$, these integrals can be computed as
\ba
\int_{\mathbb R^d} \ud Q \, \prod_{i=1}^3\frac{\Gamma(\Delta_i)}{(-2P_i\cdot Q)^{\Delta_i}}=\pi^{\frac d2}\left(\prod_{i<j}^3\frac{1}{(-2P_i\cdot P_j)^{\delta_{ij}}}\right) I(x)\,, \qquad x\equiv \frac{-2 P_{12}}{(-2P_{13})(-2P_{23})}
\ea
with
\ba
I(x)=\int_{-i\infty}^{i\infty} \frac{\ud c}{2\pi i} \Gamma(-c)\Gamma(\delta_{12}-c)\Gamma(\delta_{13}+c)\Gamma(\delta_{23}+c) x^c \ .
\ea
Finally, closing the contour and picking up the poles on the right-hand side, this is determined to be
\begin{multline}
I(x)=
\left[\prod_{i=1}^3 \Gamma\left(\mbox{$\frac d2-\Delta_i$}\right)\right]\ _2 F_1\left(\mbox{$\frac d2-\D_1,\frac d2-\D_2,1+\D_3-\frac d2$},x\right)\\+\Gamma(\Delta_1)\Gamma(\Delta_2)\Gamma(\mbox{$\Delta_3-\frac d2$}) z^{\frac d2-\D_3}\ _2 F_1\left(\mbox{$\D_1,\D_2,1-\D_3+\frac d2$},x\right) \ .
\end{multline}

\section{$\Phi^4$ interactions \la{supp}}
This is an appendix to Section \ref{P4} where we provide some explicit formulae that were omitted from the main text.

The coefficients $d_n$ of the blocks $G_{2\tilde \D+2n}$ in the expansion \rf{bb4} are, to order $O(g)$, a product of explicitly known quantities,
\be
d_n = \mu_{2\tilde \D+2n}^{\tilde \Phi^2} \, \l_{2\tilde \D+2n}^{\D\D} = \left(\mu_{2\tilde \D+2n}^{\tilde \Phi^2} \l_{2 \tilde\D+2n}^{\tilde\D \tilde\D}\right) \left(\frac{\l_{2\tilde \D+2n}^{\D\D}}{\l_{2 \tilde\D+2n}^{\tilde\D \tilde\D}}\right) \ .\la{pf}
\ee
To leading order, $\mu_{2\tilde \D+2n}^{\tilde \Phi^2} \l_{2 \tilde\D+2n}^{\tilde\D \tilde\D}$ are the coefficients \rf{coef} appearing in the boundary block expansion of the form factor $\langle \Phi^2 \, \phi \, \phi \rangle$ in GFF theory,
\be
\mu_{2\tilde \D+2n}^{\tilde \Phi^2} \l_{2 \tilde\D+2n}^{\tilde\D \tilde\D} =  \frac{(-1)^{n}}{n!} \, \frac{(\Df)_n^2}{(2\Df-\frac d2+ n)_n} +  O(g)  \la{B1}
\ee
while the ratio of structure constants may be extracted from, e.g., ref.\ \cite{Hijano:2015zsa},
\begin{align}
&\frac{\l_{2\tilde \D+2n}^{\D\D}}{\l_{2 \tilde\D+2n}^{\tilde\D \tilde\D}} = g \frac{(-1)^{n+1} 2^{-4-d}\pi^{-d/2} \G(\tilde \D)^2}{\G(\tfrac d2) \G(\D)^2} \la{B2}\\
&\qquad \qquad  \times\frac{\G(\D-\tilde \D-n)\G(\tfrac{d}{2}+n)\G(\tilde \D+n) \G(\tilde \D+n+\ha- \tfrac d2)\G(\D+\tilde \D+n-\tfrac d2)}{\G(\tilde \D+n+\ha)\G(\tilde \D+n+1-\tfrac d2)\G(2\tilde \D+n+1-d)} + O(g^2) \no \ .
\end{align}
The product of \rf{B1} and \rf{B2} gives the coefficients $d_n$ in \rf{pf},
\begin{align}
&d_n =  \frac{2^{-4-d} \pi^{-d/2}\G(\tfrac{d}{2}+n)}{n! \, \G(\tfrac{d}{2})} \la{7sol}\\
&\qquad \times \frac{\G(\D_0-\D_1-n)\G(\D_1+n)^3 \G(\tfrac{1}{2}-\tfrac{d}{2}+\D_1+n)\G(-\tfrac{d}{2}+\D_0+\D_1+n)\G(-\tfrac{d}{2}+2\D_1+n)}{\G(\D_0)^2\G(\ha + \D_1+n)\G(1-\tfrac{d}{2}+\D_1+n)\G(1-d+2\D_1+n)\G(-\tfrac{d}{2}+2\D_1+2n)}\no \ .
\end{align}
The other tower of coefficients $c_m$ (for the blocks $G_{2\D+2m}$) are
\begin{align}
&c_m = \frac{ (-1)^{m}\,  \G(m+\D_0)^2\G(2-\tfrac{d}{2}+2\D_0)}{m! \,  \G(\D_0)^2 \G(2m-\tfrac{d}{2}+2\D_0)}  \,    c(\D_0+m) \ , \la{5sol}\\
&c(\D) := \frac{2^{-2(2+\D_1)} (d-4\D_1) \pi^{\ha(1-d)}\G(\D_1)\G(-\tfrac{d}{2}+2\D_1)}{(d-4\D)\G(\tfrac{1}{2}+\D_1)\G(1-\tfrac{d}{2}+\D_1)^2} \la{6sol} \\
&  \qquad  \times \Big[ \frac{d\D_1}{(1-\D+\D_1)(1+2\D_1)(\D_1+1-\tfrac{d}{2})}  \no
\\ \no
&  \qquad  \qquad \qquad \qquad \qquad \times{}_5 F_4 \left(\begin{matrix}1+\tfrac{d}{2}, 1+\D_1,\tfrac{3}{2} - \tfrac d2 + \D_1,1-\D+\D_1,1-\tfrac d2+2\D_1  \\
\tfrac 32+\D_1,2-\tfrac d2+\D_1, 2-\D+\D_1,2-d+2\D_1\end{matrix}, 1 \right)\\
& \qquad \qquad   + \frac 1{\D_1-\D} {}_5 F_4 \left(\begin{matrix}\tfrac{d}{2}, \D_1,\tfrac{1}{2} - \tfrac d2 + \D_1,-\D+\D_1,-\tfrac d2+2\D_1 \\
\tfrac 12+\D_1,1-\tfrac d2+\D_1, 1-\D+\D_1,1-d+2\D_1\end{matrix}, 1 \right)\Big] + (\D \to \tfrac d2 - \D) \ . \no
\end{align}

We will also use the fact that the product of two bulk-to-bulk propagators between the same points has the following expansion \cite{Fitzpatrick:2011hu}
\begin{align}
&P^{BB}_{\D_a}(X',X) P^{BB}_{\D_b}(X',X) = \sum_{n=0}^\infty a_n \,  P^{BB}_{\D_a+\D_b+2n}(X',X)  \ , \\
&a_n = \frac{\pi^{-d/2}(2\D_a+2\D_b+4n-d) \G(d/2+n)}{n! \, \G(d/2)} \\
&\qquad \times \frac{\G(\D_a+n)\G(\D_b+n)\G(\D_a+\D_b+n-d/2)\G(\D_a+\D_b+2n+1-d)}{ \G(\D_a+n+1-d/2)  \G(\D_b+n+1-d/2)  \G(\D_a+\D_b+n+1-d)  \G(\D_a +\D_b +2n)} \ . \no
\end{align}

\section{General flat space limit} 
\label{app:flat}
In this appendix to Section \ref{sec:flat}, we present results relevant for determining the flat space limit of boundary blocks for general $d$, $\Delta_1,\Delta_2$. We want to determine the flat space limit of $\tilde G_{\Delta}^{12}$. The idea is to start from the Casimir equation satisfied by $G_{\Delta}^{12}$,
\ba
\left[\tilde C^{2}_z-\Delta(\Delta-d)\right] \tilde G_{\Delta}^{12}(z)=0 \label{eq:c2t}
\ea
with $\tilde C^{2}_z$ given by
\ba
\tilde C^{12}_z f(z) = (1-z)^{\frac{\Delta_{12}}2} C^2_{\frac z{z-1}}[(1-z)^{-\frac{\Delta_{12}}2} f(z)]\ .
\ea
We now plug in an ansatz
\ba
\tilde G_{\Delta}^{12}(z)=\frac{[N(z)]^{\frac{\Delta}2}}{D(z)}
\ea
and expand \reef{eq:c2t} for large $\Delta$ with $\hat \Delta_{12}\equiv \Delta_{12}/\Delta$ fixed. Setting
\ba
r(z)=\frac{1-\sqrt{1-z}}{1+\sqrt{1-z}} \ .
\ea
we can find
\ba
N(z)&=\frac{4 \left(\frac{1-\hat \Delta_{12}}{1+\hat \Delta_{12}}\right)^{\hat \Delta_{12}} (1-z)^{-\hat \Delta_{12}} \left(\frac{\nu
   (r)+\hat \Delta_{12} (1+r)}{\nu (r)-\hat \Delta_{12}(1+r)}\right)^{\hat \Delta_{12}}}{\left(1-\hat \Delta_{12}^2\right) \sqrt{\frac{1+\nu (r)+r \left(2 \Delta
   _{12}^2+\nu (r)+r\right)}{1-\nu(r)+r \left(2 \hat \Delta_{12}^2-\nu (r)+r\right)}}}\,,\\
   D(z)&=\frac{2^{d/4} (1+r)^{\frac{d-1}2}\sqrt{\nu (r)}}{ \left(1+r^2+2 \hat\Delta _{12}^2 r+(1+r) \nu (r)\right)^{d/4}}\,,\\
   \nu(r)&=\sqrt{(1-r)^2+4 \Delta _{12}^2 r}\ ,
\ea
where the normalisation is fixed by the small $z$ asymptotics.
Let us introduce now the BOE density
\ba
\hat c_{\Delta}^{\mbox{\tiny free}}=\frac{\Gamma \left(\frac{ \Delta +\Delta _1-\Delta _2}2\right) \Gamma \left(\frac{\Delta
   -\Delta _1+\Delta _2}2\right) \Gamma \left(\frac{-d+\Delta +\Delta _1+\Delta
   _2}2\right)}{\Gamma \left(\Delta _1\right) \Gamma \left(\frac{ \Delta -\Delta _1-\Delta
   _2+2}2\right) \Gamma \left(\Delta _2\right) \Gamma \left(\Delta -\frac{d}{2}\right)} \ .
\ea
Then in the same large $\D$ limit studied in Section \ref{sec:flat}, with $z<0$, one finds from the above expressions
\ba
\hat c_{\Delta}^{\mbox{\tiny free}}\frac{(1-z)^{-\Delta_{12}}}{(-z)^{2\Df}} \tilde G_{\Delta}^{12}(\mbox{$\frac{z}{z-1}$})\sim 2 \mathcal N_{\D_1,\D_2}(\Delta,s_z) \ ,
\ea
where
\ba
\frac{s_z}4=1-\frac{4 \D_1 \D_2}{(\D_1+\D_2)^2}\, z
\ea
and
\ba
\mathcal N_{\D_1,\D_2}(\Delta,s)&=\sqrt{\frac{\sigma}\pi} \exp\left[-\sigma \left(\Delta-\frac {\D_1+\D_2}2 \sqrt{s}\right)^2\right]\,,\\
\Big( \sigma&=\frac{8 \Delta _1 \Delta _2 s}{\left(\Delta _1+\Delta _2\right) (s-4) \left[\left(\Delta _1+\Delta _2\right)^2 s-4 \Delta
   _{12}^2\right]} \Big)
\ea
is a normalised Gaussian centered at $s=4[\frac{\Delta}{\D_1+\D_2}]^2$, tending to a delta function in the large $\D$ limit.

%%%%%%%%%%%%%%%%%%%%%%%%%%%%%%%%%%%%%
%%%%%%%%%%%%%%%%%%%%%%%%%%%%%%%%%%%%%
\bibliography{bib}
\bibliographystyle{JHEP}

	\end{document}